\documentclass[sigconf]{acmart}
\usepackage{longtable}
\usepackage{booktabs}
\usepackage{pgfplots}
\pgfplotsset{compat=1.18}
\usepackage{tikz}
\usepackage{graphicx}
\usepackage{subcaption} %
\usepackage{caption}
\usepackage{xcolor} %
\usepackage{enumitem}
\usepackage{url}
\usepackage{listings}

\newcommand{\graybox}[1]{%
  \medskip
  \noindent
  \fcolorbox{black}{black!10}{%
    \begin{minipage}{\dimexpr\linewidth-2\fboxsep-2\fboxrule}%
      \small #1%
    \end{minipage}%
  }%
  \medskip
}

\AtBeginDocument{%
  }

\newcommand{\deleted}[1]{\ignorespaces}
\newcommand{\add}[1]{{\color{black}{#1}}}

\usepackage{xspace}

\PassOptionsToPackage{svgnames}{xcolor}
\usepackage{xspace, xpunctuate}
\usepackage{enumitem}  %
\usepackage{multirow}
\definecolor{keywordcolor}{rgb}{0.56, 0.13, 0.00}
\definecolor{ndkeywordcolor}{rgb}{0.05, 0.46, 0.17}
\definecolor{commentcolor}{rgb}{0.41, 0.64, 0.70}
\definecolor{stringcolor}{rgb}{0.25, 0.44, 0.63}
\lstdefinelanguage{TypeScript}{
  keywords={typeof, new, true, false, catch, function, return, null, catch, switch, var, if, in, while, do, else, case, break, boolean},
  morekeywords={[2]{class, export, throw, implements, import, this}},
  identifierstyle=\color{black},
  sensitive=false,
  comment=[l]{//},
  morecomment=[s]{/*}{*/},
  commentstyle=\color{commentcolor}\ttfamily,
  stringstyle=\color{stringcolor}\ttfamily,
  morestring=[b]',
  morestring=[b]"
}
\colorlet{punct}{red!60!black}
\definecolor{delim}{RGB}{20,105,176}
\colorlet{numb}{magenta!60!black}

\lstdefinelanguage{json}{
    basicstyle=\footnotesize\ttfamily,
    numbers=left,
    numberstyle=\color{gray}\footnotesize\ttfamily,
    numbersep=4pt,
    tabsize=2,
    showstringspaces=false,
    breaklines=true,
    literate=
     *{0}{{{\color{numb}0}}}{1}
      {1}{{{\color{numb}1}}}{1}
      {2}{{{\color{numb}2}}}{1}
      {3}{{{\color{numb}3}}}{1}
      {4}{{{\color{numb}4}}}{1}
      {5}{{{\color{numb}5}}}{1}
      {6}{{{\color{numb}6}}}{1}
      {7}{{{\color{numb}7}}}{1}
      {8}{{{\color{numb}8}}}{1}
      {9}{{{\color{numb}9}}}{1}
      {:}{{{\color{punct}{:}}}}{1}
      {,}{{{\color{punct}{,}}}}{1}
      {\{}{{{\color{delim}{\{}}}}{1}
      {\}}{{{\color{delim}{\}}}}}{1}
      {[}{{{\color{delim}{[}}}}{1}
      {]}{{{\color{delim}{]}}}}{1},
}

\lstdefinelanguage{mylang}{
}
\lstdefinestyle{mystyle}{
  basicstyle=\footnotesize\ttfamily,
  numbers=none,
  frame=none,
  columns=flexible,
  xleftmargin=0pt,
  aboveskip=0pt,
  belowskip=0pt,
  language=mylang
}

\setcopyright{acmlicensed}
\copyrightyear{2025}
\acmYear{2025}
\acmDOI{10.1145/3746059.3747719}
\acmConference[UIST '25]{The 38th Annual ACM Symposium on User Interface Software and Technology}{September 28-October 1, 2025}{Busan, Republic of Korea}
\acmISBN{979-8-4007-2037-6/2025/09}

\acmSubmissionID{4408}

\begin{document}

\title{InReAcTable: LLM-Powered Interactive Visual Data Story Construction from Tabular Data}

\author{Gerile Aodeng}
\orcid{0009-0008-8661-8931}
\affiliation{%
  \institution{Beijing Institute of Technology}
  \city{Beijing}
  \country{China}
}
\email{gerile@bit.edu.cn}

\author{Guozheng Li}
\authornote{indicates the corresponding author.}
\affiliation{%
  \institution{Beijing Institute of Technology}
  \city{Beijing}
  \country{China}
}
\email{guozheng.li@bit.edu.cn}

\author{Yunshan Feng}
\affiliation{%
  \institution{Beijing Institute of Technology}
  \city{Beijing}
  \country{China}
}
\email{yunshanfeng@bit.edu.cn}

\author{Qiyang Chen}
\affiliation{%
  \institution{Beijing Institute of Technology}
  \city{Beijing}
  \country{China}
}
\email{qiyangchen@bit.edu.cn}

\author{Yu Zhang}
\orcid{0000-0002-9035-0463}
\affiliation{%
  \institution{University of Oxford}
  \city{Oxford}
  \country{United Kingdom}
}
\email{yuzhang94@outlook.com}

\author{Chi Harold Liu}
\affiliation{%
  \institution{Beijing Institute of Technology}
  \city{Beijing}
  \country{China}
}
\email{chiliu@bit.edu.cn}

\renewcommand{\shortauthors}{Gerile et al.}

\begin{abstract}
Insights in tabular data capture valuable patterns that help analysts understand critical information.
Organizing related insights into visual data stories is crucial for in-depth analysis.
However, constructing such stories is challenging because of the complexity of the inherent relations between extracted insights. 
Users face difficulty sifting through a vast number of discrete insights to integrate specific ones into a unified narrative that meets their analytical goals. 
Existing methods either heavily rely on user expertise, making the process inefficient, or employ automated approaches that cannot fully capture their evolving goals. 
In this paper, we introduce InReAcTable, a framework that enhances visual data story construction by establishing both structural and semantic connections between data insights. 
Each user interaction triggers the Acting module, which utilizes an insight graph for structural filtering to narrow the search space, 
followed by the Reasoning module using the retrieval-augmented generation method based on large language models for semantic filtering, ultimately providing insight recommendations aligned with the user’s analytical intent.
Based on the InReAcTable framework, we develop an interactive prototype system that guides users to construct visual data stories aligned with their analytical requirements. 
We conducted a case study and a user experiment to demonstrate the utility and effectiveness of the InReAcTable framework and the prototype system for interactively building visual data stories.

\end{abstract}



\maketitle

\section{Introduction}
\label{sec:introduction}

Tabular data is a fundamental format for representing complex information in a wide range of domains, from business analytics and scientific research to social sciences and healthcare~\cite {10.1007/s00778-021-00714-0, 2018-Expandable-Dou, 2013-Automatic-Chen, 2010-interactive-ward, li2023HiTailor, li2024InsigHTable, Ghosh2018VI}.
The tabular data consists of many data insights that are valuable patterns and highlight key statistical information, such as trends and outliers~\cite{ding_2019_quickinsights}. 
Visual data storytelling involves organizing related data insights into a coherent narrative, using visualizations to convey users' complex analysis findings in an easily understandable way~\cite{6064988, DBLP:journals/cga/LeeRIC15}. 
\add{Within the visual data storytelling process, visual data story construction is a critical and challenging task, which requires users to have a deep understanding of the dataset,}
\deleted{Although visual data stories are beneficial for data analysis, their construction is a challenging task, as it requires users to have a deep understanding of the dataset,} efficiently select relevant data insights that match their analytical goals and logically connect these insights, making the process highly dependent on the individual expertise and experience of the users~\cite{9903555, ma2023demonstrationinsightpilotllmempoweredautomated, weng2024insightlensaugmentingllmpowereddata, shi_2021_calliope}.

The above difficulties have motivated many studies on visual data story construction.
Some studies~\cite{tang_2017_extracting, ding_2019_quickinsights} have streamlined the data insight extraction process, using rule-based algorithms to calculate insights from tabular data. 
With extracted insights, different approaches have been developed to assemble these elements into a coherent narrative.
Some methods automatically generate data stories based on predefined rules~\cite{wang_2020_datashot, shi_2021_calliope, 8973383}.
Although automated approaches can rapidly produce narratives, the results may not align with users' analytical goals.
Other methods support interactive construction by allowing user participation through inputs~\cite{ma2023demonstrationinsightpilotllmempoweredautomated} and refinements~\cite{9903555, liCoInsightVisualStorytelling2024, Socrates2024Wu}.
Although interactive methods offer greater control, they tend to be inefficient and are based on trial and error. 
Therefore, a method to efficiently construct visual data stories is desirable, which takes users' specific needs into account and ensures that the constructed narrative aligns with the user's analytical goals.

However, designing such a technique to construct coherent visual data stories faces two primary challenges.
First, tabular data extraction typically results in a large number of discrete insights, leaving users overwhelmed by the sheer volume and complexity of independent insights. 
Users are required to select a subset of insights that aligns with their specific analytical goals from a huge search space, which requires significant expertise and experience, making the construction of coherent data narratives particularly challenging.
Second, the relevance of insights is shaped by the evolving analytical intent of users---an insight that is highly relevant in one context may be peripheral in another. 
Therefore, dynamically associating insights based on user intent remains a difficult problem in data narrative construction.

To address these challenges,
we propose the InReAcTable framework, a paradigm that guides users in constructing visual data stories for tabular data. 
Each user \underline{\textbf{In}}teraction triggers an iterative loop in which the \underline{\textbf{Re}}asoning and \underline{\textbf{Act}}ing modules collaborate.
The Acting module constructs a subspace graph and enables structural filtering to narrow the search space.
The Reasoning module uses the retrieval-augmented generation (RAG) method to obtain the top‑$k$ relevant insights, which are subsequently input into large language models (LLMs) for semantic reasoning, ultimately recommending data insights that align with the analytical intent of the users.
This integrated framework streamlines the data story construction workflow and improves narrative coherence.
Based on the InReAcTable framework, we develop a prototype system to assist users in exploring tabular data and constructing visual data stories. 
Within the system, users can pose natural language questions guided by insight visualizations and make interactive selections based on recommendations by the system, while maintaining flexibility to refine and revise their analysis.

We evaluate the InReAcTable framework and the prototype system through a real-world use case and a user experiment. 
The case study with a domain analyst demonstrates the practicality of InReAcTable in enhancing the ability of users to build coherent and insightful data stories. 
In addition, we compare InReAcTable with two widely adopted technologies: AWS QuickSight~\cite{quick_sight}, representing automated generation tools, and vizGPT~\cite{vizgpt}, representing LLM-driven tools, to ensure a comparison between two distinct categories of systems. 
Participants were tasked with building data stories under freeform exploration and target-driven exploration. 
We employed a combination of quantitative and qualitative evaluations, along with user feedback, to thoroughly evaluate the system's performance. 
The results indicate that InReAcTable significantly enhances the effectiveness of data exploration and story construction, and received positive feedback regarding both the design of the system and the overall user experience.

In summary, the main contributions of this paper are as follows: 
\begin{itemize}[leftmargin=*]
\item We propose InReAcTable, an LLM-powered framework that enhances data story generation by enabling users to perform structural and semantic filtering to find related data insights, facilitating coherent narrative construction. 

\item Based on the InReAcTable framework, we develop an interactive system that guides users to construct data stories aligned with their analytic requirements.

\item We provide a case study and a user experiment to demonstrate the utility and usability of InReAcTable for interactively visual data story construction.
\end{itemize}

The source code for InReAcTable is available at \url{https://github.com/bitvis2021/InReAcTable}.

\section{Related Work}
\label{sec:related-work}

This section reviews the literature on the visual data story construction for tabular data and the applications of LLMs in visualizations to position our work. 

\subsection{Visual Data Story Construction}

Researchers have developed various frameworks and techniques to automate the extraction and organization of data insights. 
Tang et al.~\cite{tang_2017_extracting} propose the concept of data insight in a comprehensive way and enable users to extract the top-$k$ insights from multidimensional data utilizing a scoring function.
Furthermore, QuickInsight~\cite{ding_2019_quickinsights} refines this idea by providing a unified abstract definition that integrates various types of data patterns into a single framework, allowing automatic extraction of information from tabular data.
However, the automatically extracted insights are independent and scattered, which is insufficient to support complex data analysis tasks. 
Analysts still need to sift through and interpret the vast amount of insights generated.

Creating a coherent data story requires connecting discrete data insights with specific relations. 
We divide data story generation techniques into two categories. 
The first category emphasizes the automatic generation of data stories.
For example, DataShot~\cite{wang_2020_datashot} organizes the generated data facts into different topics, with each topic creating an infographic to communicate complex data intuitively.
Calliope~\cite{shi_2021_calliope} further defines six types of logical relationships between insights, organizing them in a logical sequence to automatically form a coherent narrative.
MetaInsight~\cite{maMetaInsightAutomaticDiscovery2021} generates structured semantic insights by forming inductive hypotheses and conducting validity inquiries within homogeneous data scopes. 

The second category involves user participation in the data story construction process and can be further divided into three types. 

\begin{itemize}[leftmargin=*]
\item The first type helps users link data insights and build visual data stories by providing visual cues. 
For example, CoInsight~\cite{liCoInsightVisualStorytelling2024} models the header structure of hierarchical tables, allowing users to connect insights across different data scopes. 

\item The second type focuses on recommending data insights to reduce the burden of manually browsing large volumes of information.
For example, ChartStory~\cite{zhao2023} allows users to input pre-generated charts and provides recommendations for partitions, layouts, and captions for narrative construction.
Shi et al.~\cite{8973383} propose a reinforcement learning method to suggest the chart sorting process to provide choices for user design.
Erato~\cite{9903555} introduces an interpolation algorithm to smooth user-provided keyframes to create a coherent data story.
\add{To further improve the flexibility, NL4DV~\cite{DBLP:journals/tvcg/NarechaniaSS21} provides a toolkit for mapping natural language queries to analytic specifications and recommending corresponding data visualizations.}

\item The third type enables users to refine the generated visual data stories. 
For example, 
\add{Groot~\cite{DBLP:conf/visualization/GathaniCSS24} allows users to configure chart elements directly and receive insight recommendations based on their manipulations, offering flexibility for interactively editing insights.}
Socrates~\cite{Socrates2024Wu} incorporates user feedback into the story generation process, allowing iterative refinement.
\end{itemize}

Although there are extensive studies on tabular data story construction, they predominantly rely on static rules to connect insights, limiting their ability to support complex analytical tasks.
Our approach focuses on constructing coherent narratives by integrating both structural and semantic relationships among insights, thereby enabling more guided and contextually enriched story construction.

\subsection{LLMs for Visualizations}
LLMs have extensive knowledge and can effectively apply the knowledge to perform various tasks based on user input~\cite{wu2023llms}, demonstrating remarkable problem-solving capabilities.
Consequently, many recent studies have taken advantage of LLMs to enhance the visual analytics process.
These studies can be broadly divided into three main categories: LLMs for visualization generation, LLMs for visual analytics, and LLMs for tabular data analysis. 

\textbf{LLMs for visualization generation.} 
LLMs are proficient in generating text with specific format~\cite{gpt4technical}. Related studies on visualization generations consist of various forms, such as imperative code and declarative languages. 
Chat2Vis~\cite{chat2vis2023} generates visualization code in Python by prompting LLMs with table schema, column types, and expressions from user queries. 
Furthermore, LIDA~\cite{LIDA2023} defines visualization generation as a four-stage generation problem and uses GPT-3.5 to generate Python code for visualization creation. 
In addition to imperative code, the JSON-based declarative grammar~\cite{jsongrammar2023}, such as Vega-Lite~\cite{vegalite2017}, also exists pervasively and is well-established, which allows users to define and encode visual mappings for the data. 
Existing studies have extensively evaluated the capability of LLMs to generate Vega-Lite specifications~\cite{li2024visualizationgenerationlargelanguage, chen2025VisEval}.
ChartGPT~\cite{chartgpt2024} adopts the least-to-most~\cite{zhou2023leasttomostpromptingenablescomplex} principle to decompose the visualization generation task and then solve it sequentially. 
In contrast to our work, the above studies focus primarily on the code generation capability of LLMs.

\textbf{LLMs for visual analytics.}
Data analysts often need expertise in data visualization principles and domain knowledge to use these systems effectively~\cite{hutchinson2024llm}. 
Recent studies try to address these challenges and improve the visual analytics process through LLMs.
GPT4-Analyst~\cite{Cheng2023DataAnalysts} uses prompts to direct GPT-4~\cite{gpt4technical} in visual analytics tasks. 
In this thread, LEVA~\cite{leva2024} integrates LLMs into visual analytics workflows at three critical stages: onboarding, exploration, and summarization.
LightVA~\cite{zhao2024lightvalightweightvisualanalytics} further advances the field by enabling automated task planning and execution through a lightweight recursive agent framework.
However, directly applying LLMs to visual data analysis presents several challenges, such as biases in the analysis, limited reasoning capabilities~\cite{lingoRoleChatGPTDemocratizing2023}, difficulty in comprehending the global context and ensuring adaptability across diverse datasets~\cite{zhaTableGPTUnifyingTables2023}. 
The opacity in the decision-making process is another concern that affects the effectiveness and trustworthiness of LLMs in these applications~\cite{zhangLargeLanguageModels2023}. 
These works primarily focus on integrating LLMs within visual analytics frameworks to enhance system usability rather than targeting explicit data analysis tasks.

\textbf{LLMs for tabular data analysis.}
Recent works have developed LLM-based tools for tabular data analysis.
Some approaches focus on leveraging LLMs to generate textual reports that encapsulate insights derived from data.
LLM4Vis~\cite{LLM4Vis2023} introduces an interpretable visualization recommendation system by guiding ChatGPT to produce detailed text descriptions for tabular datasets. 
InsightPilot~\cite{ma2023demonstrationinsightpilotllmempoweredautomated} treats LLMs as an autonomous agent to perform end-to-end data exploration and generate narrative reports. 
DataTales~\cite{sultanum2023datatalesinvestigatinguselarge} leverages LLMs to generate textual stories from an input chart, with the aim of producing editable articles without visualizations.

In parallel, the other research thread focuses on producing visualizations with explanations through interactive dialogue.
VizGPT~\cite{vizgpt} translates natural language queries into visualizations and supports iterative refinement through conversational interaction. 
AI Threads~\cite{hong2023conversationalaithreadsvisualizing} introduces a dialogue structure with multi-threads to better mimic human analytical reasoning. 
Islam et al.~\cite{islam2024datanarrativeautomateddatadrivenstorytelling} and Weng et al.~\cite{weng2024insightlensaugmentingllmpowereddata} 
further enhance this process by employing LLM agents to simulate different roles in the data analysis workflow.

Although these end-to-end methods effectively address individual questions, they typically do not explicitly construct or model the relationships between different insights. 
This limitation results in fragmented interpretations, making it difficult to build a cohesive understanding of the data and potentially overlooking underlying patterns. 
Our approach addresses this gap by combining the reasoning capabilities of LLMs with the searching ability from insight graphs of computations, alongside the user's knowledge and experience, to enable efficient and effective visual data story construction.

\section{Preliminary Study}
\label{sec:preliminary_study}

We conducted a preliminary study to clarify the current practice for visual data story construction of tabular data. 
This study also aimed to establish design requirements for the target system. 

\subsection{Study Design}

\subsubsection{Participants}
We conducted unstructured interviews with four experts with extensive data analysis and visualization experience. These experts include
a professor engaged in data visualization and data storytelling (E1, age 45), two data scientists from a well-known financial and insurance company (E2, age 34, and E3, age 28), 
and a researcher focused on artificial intelligence for visualization (E4, age 26).
Each expert has spent more than 5 years deeply involved in the field of visual analytics or tabular data analysis.

\subsubsection{Procedure}
To better understand experts' approaches to visual data story construction in tabular data, we conducted individual interviews lasting between 1 and 1.5 hours. 
Initially, experts demonstrated the tools they use routinely to extract data insights and the process of building visual data stories.
They then provided detailed walk-throughs of their process, using think-aloud protocols to articulate each analysis step. 
Following these demonstrations, experts were encouraged to reflect on the strengths and weaknesses of the tools they used. 

To further explore their requirements, we asked three key questions:
(1) What are the key procedures in the visual data story construction from tabular data? 
(2) What are the most challenging steps in the process? 
(3) What features and functions should an ideal system for constructing visual data stories from tabular data have? 
After the interviews, we summarized and analyzed the feedback of the experts, which guided the design of our interaction framework.

\add{Our interviews with four domain experts revealed several recurring patterns. 
Experts described a typical workflow that begins with an initial broad exploration to familiarize themselves with the dataset, followed by a more focused stage of organizing selected insights into coherent narratives.
They commonly cited challenges such as locating meaningful insights, maintaining analytical focus, and the cognitive effort required to connect disparate pieces of information into a logical story structure.
These recurring breakdowns and frustrations, along with key reflections on system needs, helped shape our design requirements. }

\subsection{Design Requirements}
\label{sec:design_requirements}

Informed by the interview results and prior literature, we distilled four design requirements for an interactive visual data story construction framework:

\begin{itemize}[leftmargin=22pt]
    \item[\textbf{DR1}] \textbf{Support automated insight extraction\deleted{ and expressive visualizations}. }
    Manually extracting insights from large datasets is often repetitive and time-consuming, which can overwhelm users and hinder their ability to focus on higher-level analysis. 
    \add{As E2 commented, ``\textit{It often takes me hours just to sift through pivot tables or metrics to find something that looks interesting. It’s hard to know what to focus on.}''
    Similarly, E3 emphasized that ``\textit{When the dataset is complicated, it’s hard to tell whether I’ve missed something important unless I check every dimension manually.}''
    These frustrations highlight the need for automated insight extraction that reduces manual effort and ensures important patterns are not overlooked~\cite{wang_2020_datashot, liCoInsightVisualStorytelling2024, li2024insightableinsightdrivenhierarchicaltable}.
    }
    \deleted{Therefore, the framework should enable automated data insight extraction. } 
    \deleted{In addition, to reduce the cognitive load on users and improve their understanding, it is critical to implement concise and expressive visualizations to express data insights~\cite{shi_2021_calliope,DBLP:journals/corr/abs-2008-11015}. }

    \item[\textbf{DR2}] \textbf{Facilitate user steering visual data story construction with evolving analytical intent.}
    Recognizing that user interests and analytical intent may shift during exploration~\cite{7192728, 10.1145/3025453.3025768}, the framework should allow user control over the data story construction process~\cite{Socrates2024Wu, li2024insightableinsightdrivenhierarchicaltable, DBLP:journals/cga/LeeRIC15, 7192728, 10.1145/3025453.3025768}.
    \add{E4 noted that ``\textit{I wish tools could pick up on what I’m trying to do and suggest other paths, like different dimensions or complementary insights,}''}
    which highlights the need for interactive steering to keep generated narratives aligned with evolving user requirements, rather than relying on static and predefined rules. 
    This includes dynamically recommending related insights that reflect and support users' shifting perspectives, thereby fostering an iterative visual data story construction process.

    \item[\textbf{DR3}] \textbf{Alleviate the cognitive burden \deleted{of overwhelming insights }in data story construction.}
    In real-world scenarios, tabular datasets can produce an overwhelming number of potential insights, making it challenging for users to explore the vast data space, navigate potential story fragments, and organize insights into coherent narratives~\cite{9903555}. 
    \add{As E1 described, ``\textit{I often feel like I’m drowning in insights---many are valid, but I don’t know which ones are useful for my story.}''
    E3 also shared, ``\textit{Sometimes I end up jumping between different insights without making progress, because I’m not sure which direction to follow.}''
    }
    To alleviate this burden, the framework should assist users by narrowing the search space to insights that are most pertinent to the user’s current analytical context, thereby supporting efficient and focused data story construction.
    
    \add{Moreover, experts mentioned that during analysis, they primarily relied on visual representations to obtain information rather than inspecting raw data directly. 
    This highlights the importance of concise and expressive visualizations in improving user understanding and reducing cognitive effort when interpreting data insights~\cite{shi_2021_calliope,DBLP:journals/corr/abs-2008-11015}.}

    \item[\textbf{DR4}] \textbf{Support different stages of visual data story construction.}
    Interviews reveal that data story construction generally involves two primary stages.
    In the initial stage, users often lack prior knowledge of the dataset and need to explore a wide range of insights to establish their analytical goals~\cite{DBLP:journals/tvcg/LiY23}.
    As E1 remarked, ``\textit{At the beginning, I was often unfamiliar with the dataset and felt overwhelmed by the large number of extracted insights.}''
    As the analysis progresses, the focus shifts to a targeted discovery of insights aligned with the refined goals.
    As E3 mentioned, ``\textit{Once I have a clear exploration goal, I need to find potentially relevant insights and validate one by one whether they can be connected into a coherent data story.}" 
    To support this progression, the framework should accommodate both stages by flexibly supporting broad exploration and focused, goal-driven story construction.

\end{itemize}

\section{The InReAcTable Framework}
\label{sec:methods}

\begin{figure*}[tb]
 \centering  
    \includegraphics[width=\textwidth, 
    alt={Figure 1 shows the pipeline of the InReAcTable framework, illustrating how interaction, acting, and reasoning modules collaborate to iteratively extract, refine, and recommend insights for visual data story construction.}]
    {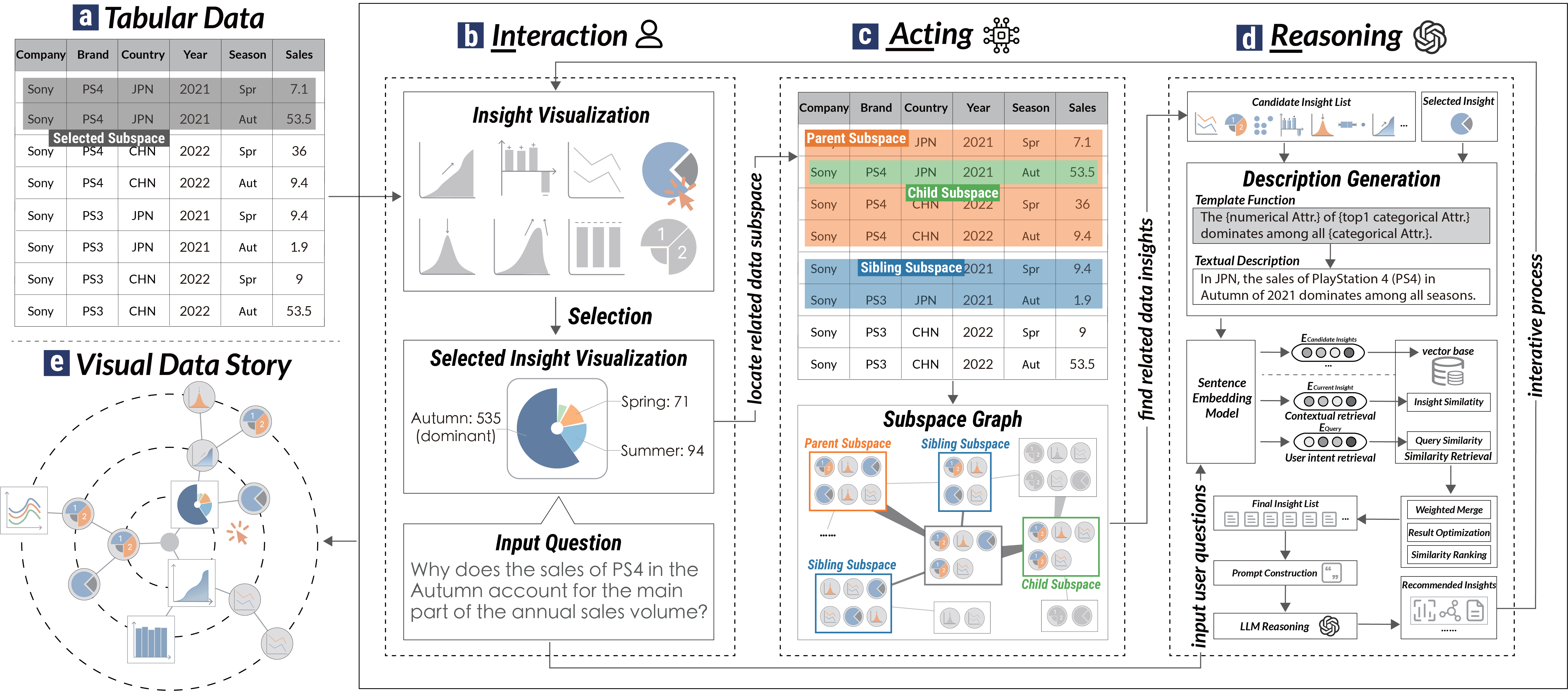}
    \caption{      
        The InReAcTable framework enables interactive visual data story construction through an iterative workflow. 
        Starting with the input of tabular data, users begin by selecting subspaces (a), which automatically triggers the insight extraction process. 
        The Interaction module (b) visualizes these insights and allows users to input questions based on the visualizations. 
        The Acting module (c) constructs a subspace graph to locate and filter relevant subspaces, generating a refined list of candidate insights through structural filtering. 
        The Reasoning module (d) converts insights into natural language descriptions and applies RAG to retrieve contextually relevant insights based on both the user’s query and the currently selected insight. 
        This process leverages the semantic understanding capabilities of LLMs to support dynamic insight recommendations. 
        The recommended insights are then fed back into the Interaction module, continuing the cycle. 
        Through this iterative process, users can review, select, and explore insights, gradually constructing a coherent and meaningful visual data story (e).}
 \label{fig:pipeline}
\end{figure*}

In this section, we first represent the abstract model of tabular data, which serves as the foundational basis for our framework.
Then, we introduce the InReAcTable framework, where each user interaction (see Sect.~\ref{sec:interaction}) is expanded into an iterative loop in which the Acting module (see Sect.~\ref{sec:acting}) and the Reasoning module (see Sect.~\ref{sec:reasoning}) collaborate, as shown in Fig.~\ref{fig:pipeline}.

\subsection{Data Model}
\label{sec:Abstract-Model}

The data model defines data subspaces, analysis entities, and data insights, which provide the foundation for insight extraction and insight graph construction.

To start, let $D := \{X_1, X_2, \ldots, X_n\}$ represents \textbf{tabular data} where each row entity contains $n$ attributes. 
These attributes can be either categorical attributes or numerical attributes.
Taking the tabular data in Fig.~\ref{fig:pipeline} as an example, a \textit{filter} operation uses an equality assertion $X = x_i$ to specify that the values for a given dimension $X$ need to match $x_i$ (e.g., Company = ``Sony'').
Applying the filter operation results in a subset of the raw table where each row entity satisfies the specified condition. 
Furthermore, we define $Locator (Loc)$ to denote the conjunction of filters across disjoint dimensions. 
For example, $Loc$(\textit{Sony}, \textit{PS4}, \textit{JPN}, \textit{2021}) specifies the conditions (\textit{Company} = ``\textit{Sony}'' AND \textit{Brand} = ``\textit{PS4}'' AND \textit{Country} = ``\textit{JPN}'' AND \textit{Year} = ``\textit{2021}''), thus defines a \textbf{data subspace}. 
For unspecified dimensions, all possible attribute values are implicitly included by default. 

An \textbf{Analysis Entity (AE)} represents a fine-grained division within a data subspace. 
Formally, an AE is defined as $AE:=\langle S, B, \\
Agg(V) \rangle$, 
where $S$ is the data subspace, $B$ is the categorical attribute selected as the dimension of breakdown, and $Agg(V)$ is the aggregate operation applied to the numerical attribute $V$ within $S$.
The entities in the row that share the same value in the breakdown dimension $B$ are aggregated to form a new entity with the value of $Agg(V)$. 
The aggregation employs mathematical functions such as $min$, $max$, and $sum$, which reduce multiple values to one.

The \textbf{data insight} represents valuable data patterns within an AE, which is defined as a 5-tuple $\langle$\textit{AE}, \textit{Type}, \textit{Category}, \textit{Score}, \textit{Description}$\rangle$. 
\textit{Type} denotes the insight type (outlier, trend, etc.) and \textit{Category} refers to its classification (point insight, shape insight, and compound insight), as detailed in Sect.~\ref{sec:interaction}.
The \textit{Score} quantifies the significance of data insight, and the \textit{Description} provides a textual representation of data insight, detailed in Sect.~\ref{sec:description}.

\subsection{Interaction Module}
\label{sec:interaction}

\begin{figure}[tb]
 \centering  
    \includegraphics[width=0.8\columnwidth, 
    alt={Figure 2 shows three categories of data insights—point, shape, and compound—each further divided into specific types with their own visualization forms.}]
    {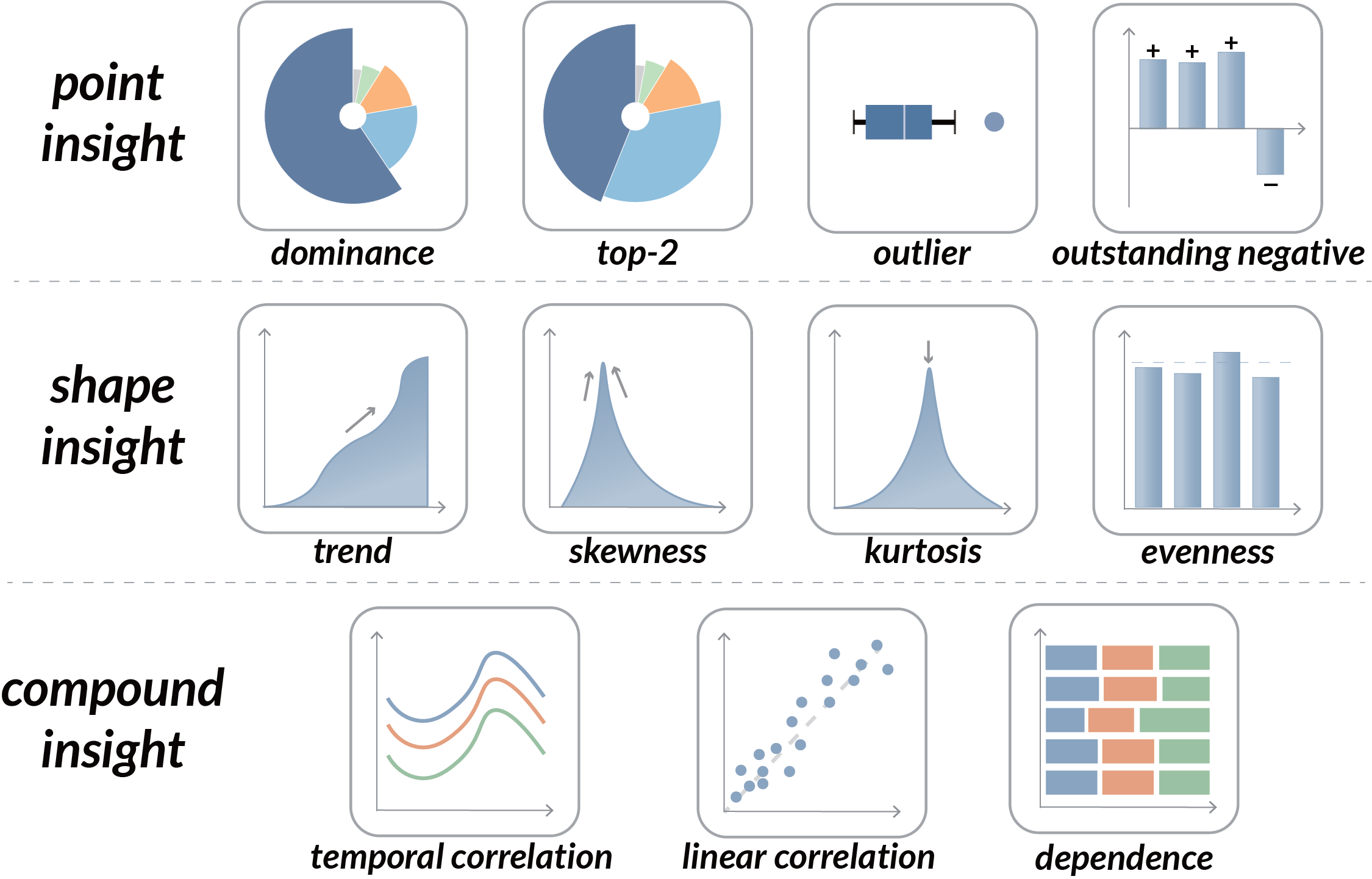}
    \caption{        
        Three categories of data insights: point insight, shape insight, and compound insight.
        They are further divided into eleven types, each corresponding to its specific visualization form.
    }
 \label{fig:insights}
\end{figure}

In the InReAcTable framework, the interaction module is designed to support user participation in the visual data story construction process. 
Specifically, at the beginning of user exploration, the interaction module extracts data insights from the user-selected subspace and presents them through appropriate visualizations, as shown in Fig.~\ref{fig:insights}, allowing users to understand underlying data patterns intuitively.
Throughout the iterative process, the interaction module takes the recommended data insights as input and transforms them into corresponding visualizations. 
With insight visualizations, the interaction module allows users to select data insights of interest, pose questions on specific visualizations, and further guide the construction of the visual data story.

To support this, we traverse all AEs, performing insight calculations and retaining only data insights with scores that meet or exceed the pre-defined thresholds. 
Based on existing works~\cite{ding_2019_quickinsights, Cheng2023DataAnalysts}, we categorize insights derived from tabular data into three main categories: point insights, shape insights, and compound insights. 
Each category represents specific characteristics and provides different analytical perspectives on the data.
We employ the definition and calculation methods of insights in previous studies~\cite{tang_2017_extracting, Cheng2023DataAnalysts}, and use the insight score and threshold in CoInsight~\cite{liCoInsightVisualStorytelling2024} to measure the significance of each insight.

\begin{itemize}[leftmargin=*]
\item Point insights focus on specific data points to highlight their significance or deviation from the norm, including dominance, top-2, outlier, and outstanding negative.
\item Shape insights focus on examining the overall distribution and structure of the data, including trend, skewness, kurtosis, and evenness.
\item Compound insights reveal complex patterns between different subgroups within an AE, including temporal correlation, linear correlation, and dependence.
\end{itemize}

To enhance the clarity and expressiveness of visual presentations, we apply mapping rules established in prior research~\cite{DBLP:journals/tvcg/MackinlayHS07, article} and follow the visualization effectiveness principles~\cite{DBLP:journals/tog/Mackinlay86}. 
Consequently, for each type of insight, we identified visualization charts to represent its data characteristics.
Visualization transforms key insights into an intuitive and accessible format, reducing the cognitive burden on users and facilitating data-driven exploration \textbf{(DR3)}.

\subsection{Acting Module}
\label{sec:acting}

When constructing visual data stories, users need to navigate a vast data space to identify insights most relevant to their analytical objectives.
However, the sheer volume of discrete insights makes direct selection and decision-making challenging.
To streamline this process, the InReAcTable framework uses the user's focused insight as a ``seed'' to trigger relevance-based exploration. 
The Acting module is designed to address this complexity by leveraging a constructed subspace graph for structural filtering to narrow the search space \textbf{(DR3)}, as shown in Fig.~\ref{fig:pipeline}(c).

\subsubsection{Subspace Graph Construction}
For subspaces in tabular data, we categorize their relations into \emph{sibling} and \emph{parent-child}, allowing the construction of a subspace graph that structurally connects subspaces.

We first define the \emph{length} of a locator as the number of its specified dimensions. For example, the length of \emph{Loc(Sony, Europe)} is two. 
For two locators with the same length and specified dimensions, but with different attribute values in only one dimension (all other dimensions having the same attribute values), we define them as a \textbf{sibling} relation.
For example, a data subspace (denoted as DS1) located at \emph{Loc(Sony, Asia)} and a subspace (denoted as DS2) located at \emph{Loc(Sony, Europe)} are siblings.
The \textbf{parent-child} relation captures the directional refinement and generalization among subspaces.
Given a locator \emph{Loc1} of length $n$ and a locator \emph{Loc2} of length $n+1$, where \emph{Loc2} matches \emph{Loc1} in its $n$ dimensions, but includes an additional dimension, we define \emph{Loc1} as the parent and \emph{Loc2} as its child.
For example, the subspace (denoted as DS3) located at \emph{Loc(Sony, Asia, Spr)} has the parent-child relation with DS1.
For any tabular data, we can construct a definitive subspace graph according to these defined relations.

\subsubsection{Structural Filtering}
The underlying subspace of an insight suggests the presence of parallel and inclusive structural relationships that can reveal interconnections between insights. 
To take advantage of this information, the Acting module enables structural filtering based on the constructed subspace graph \textbf{(DR3)}. 
First, we locate the data subspace where the currently focused insight resides and then search the graph for related subspaces with sibling and parent-child relations, with the search range adjustable via a step parameter. 
Within each subspace, aggregation operations are performed across different breakdown dimensions to partition all AEs, as defined in Sect.~\ref{sec:Abstract-Model}. 
AE represents a specific perspective on data and serves as the domain where insight extraction operations are performed.

\subsection{Reasoning Module}
\label{sec:reasoning}

The relationships between insights are dynamic, shaped by users' input questions.
Since each insight can represent multiple data characteristics, its connection to other insights depends on the context.
To address this, we introduce the Reasoning module, which leverages the semantic understanding capabilities of LLMs to assist users in constructing data stories.
As shown in Fig.~\ref{fig:pipeline}(d), the Reasoning module first transforms the insights into textual descriptions and then performs sentence embedding. 
Subsequently, similarity retrieval is executed based on both the user’s query and the currently selected insight to obtain the top‑$k$ insights. 
These insights are then input into LLMs for semantic reasoning, generating insight recommendations to the user.

\subsubsection{Description Generation}
\label{sec:description}
Processing raw tabular data is resource-intensive for LLMs, and the risk of hallucinations increases with larger input volumes, making them unreliable for further analysis ~\cite{DBLP:journals/csur/JiLFYSXIBMF23}.
To address this challenge, an efficient data format is needed to support semantic insight analysis.
High-quality textual descriptions can distill key numerical features, filtering out irrelevant and low-value information to reduce noise interference.
Therefore, candidate insights obtained through structural filtering are transformed into natural language descriptions, thereby mapping discrete attributes into a continuous semantic space.

Existing studies on automated description generation can be classified into communication-oriented and explanation-oriented approaches.
The first category~\cite{wang_2020_datashot} serves as additional annotations that provide context and background information to the graphs, with the aim of improving the communicative effectiveness of infographics in conveying insights.
The second category~\cite{DBLP:journals/tvcg/SrinivasanDES19} involves using template-based natural language generation tools (e.g., Quill\footnote{https://narrativescience.com}, Wordsmith\footnote{https://automatedinsights.com/wordsmith}) 
to convert data facts into sentences, helping users understand complex visual content~\cite{DBLP:journals/corr/abs-2307-13858} and uncovering overlooked insights~\cite{DBLP:journals/tvcg/SrinivasanDES19}.

Our target description format is designed to enable LLMs to comprehend insights without processing the entire table.
Recognizing that data patterns closely align with the type of insight, we adopt predefined templates to automatically generate descriptions tailored to each insight type.
Specifically, the insight description is defined as a 4-tuple $\langle Header, Type, Score, Description \rangle$. 
$Header$ serves as the filter condition for locating the data subspace and provides structured contextual information. 
$Type$ and $Score$ indicate the type and significance of the insight. 
$Description$ provides an intuitive explanation of the data pattern, implicitly conveying the properties of the insight (e.g., trend direction and outlier polarity), along with the breakdown dimensions and aggregation operations involved.

An example of dominance insight is shown in Fig.~\ref{fig:pipeline}(d).
In the description $\langle$Header=(\emph{JPN}, \emph{PlayStation4 (PS4)}, \emph{2021}), Type=dominance, Score=0.524, Description=``In \textbf{\emph{JPN}}, the sales of \textbf{\emph{PlayStation4(PS4)}} in \textbf{\underline{Autumn}} of \textbf{\emph{2021}} dominates among all seasons.''$\rangle$,
\emph{Header} identifies the subspace \textit{S}: \textit{Location} = ``\textbf{\textit{JPN}}'' AND \textit{Brand} = ``\textbf{\textit{PlayStation4 (PS4)}}'' AND \textit{Year} = ``\textbf{\textit{2021}}'', while the \emph{Description} indicates that a \textit{SUM} aggregation operation was performed on the \textit{Sale} attribute across the \textit{Season} dimension. 
It specifies that the \textit{dominance} attribute of this insight corresponds to \textbf{\underline{Autumn}}.

Through this template-based approach, we can efficiently generate concise and precise descriptions for each type of insight.
The description templates are available in the supplemental material.

\subsubsection{RAG-based Reasoning}
We first construct vector representations by encoding the description of each insight using the Sentence-BERT model~\cite{reimers2019sentencebertsentenceembeddingsusing}, 
generating high-dimensional semantic embeddings and storing them in a FAISS~\cite{douze2024faiss} vector database.
An approximate nearest neighbor retrieval index based on the Inverted File Index mechanism (IVF)~\cite{8733051} enables efficient similarity search at scale.
This process enables subsequent similarity computations to be performed at the semantic level, rather than relying solely on keyword-based matching.

At the retrieval stage, to balance user intent with narrative coherence, we implement a dual-path retrieval:
(1) user intent retrieval encodes the query into a vector, retrieves top-$k$ insights via cosine similarity with candidate embeddings (denoted as $C_{\text{user}}$) \textbf{(DR2)}, 
while (2) contextual retrieval performs similarity matching with users' current focused insight, retrieving $C_{\text{context}}$ \textbf{(DR3)}.
The results of both retrieval paths are combined through a weighted average:
$C_{\text {merged }}=\alpha \cdot \operatorname{Rank}\left(C_{\text {user }}\right)+(1-\alpha) \cdot \operatorname{Rank}\left(C_{\text {context }}\right)$, with $\alpha$ set to $0.7$ based on the empirical study, which favors the intent of the user.

These insights obtained through the RAG-based retrieval form a semantically related candidate set. 
The results are ranked by similarity scores and further optimized using multi-constraints from metadata, including significance filtering based on $Score$ and clustering recommendations based on $Type$ and $Category$.
The optimized subset $C_{\text {final }}=\left\{c_{1}, c_{2}, \ldots, c_{K}\right\}$ is input into LLM in the description form for reasoning.

\begin{figure}[tb]
 \centering  
    \includegraphics[width=\columnwidth]{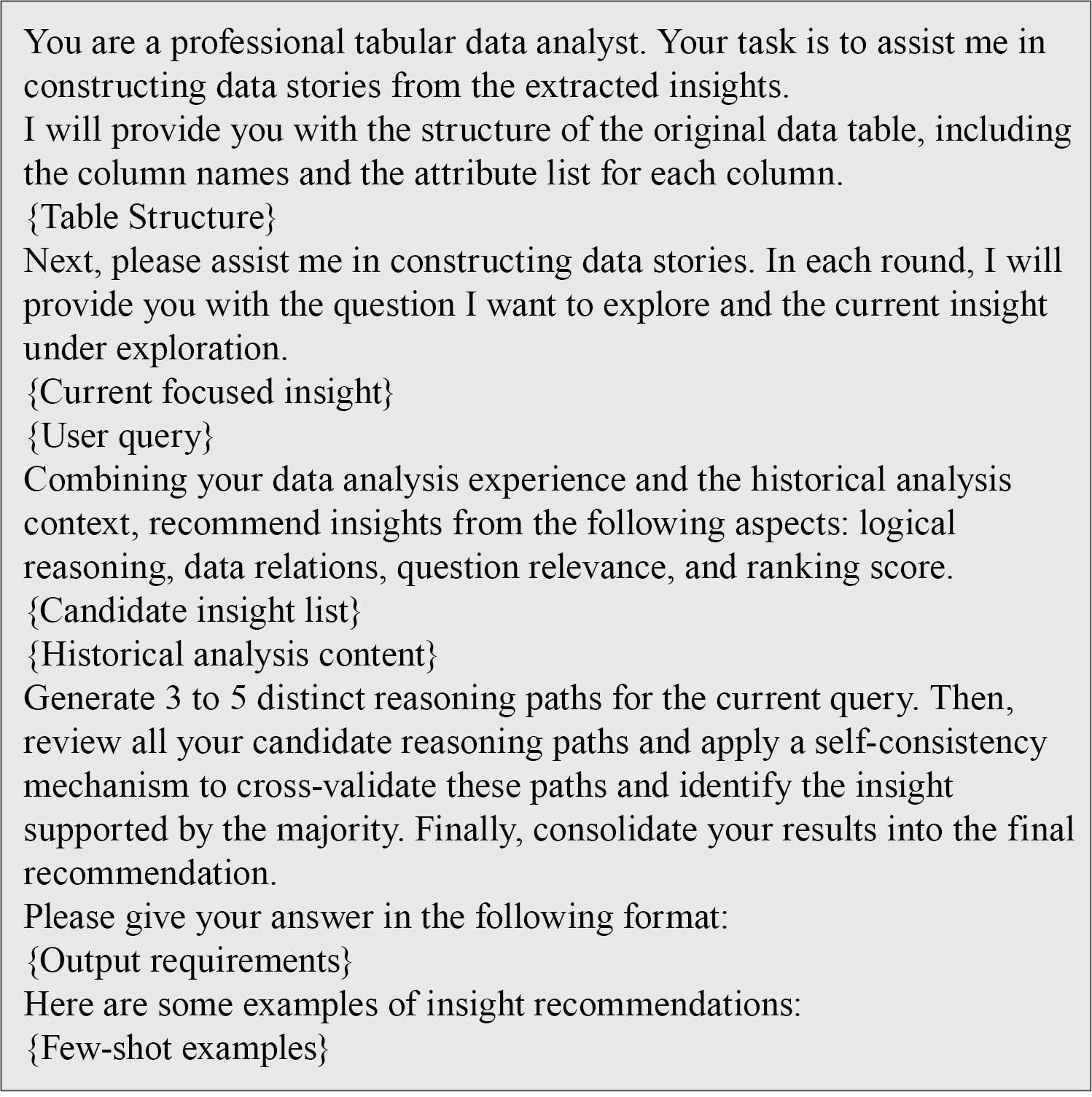}
    \caption{        
        The prompt template in the Reasoning module, designed to elicit multi-path reasoning over retrieved insights and generate recommendations.
    }
 \label{fig:prompt}
\end{figure}

We employ a Chain-of-Thought (CoT)~\cite{weiChainofThoughtPromptingElicits} prompt that uses In-Context Learning (ICL)~\cite{dong2024surveyincontextlearning} strategy to dynamically integrate users' historical analysis as context and provide few-shot examples to demonstrate the reasoning process. 
This serves as a prefix, followed by the user query, current focused insight, and candidate insight subset as input.
To enhance reasoning, we employ the Self-Consistency~\cite{wangSelfConsistencyImprovesChain2022} mechanism to sample multi-path reasoning results.
Finally, the insights recommended by LLM are returned to users in the form of visualizations and explanations to reduce cognitive load and improve user understanding \textbf{(DR1)}. The prompt design implementing these strategies is presented in Fig.~\ref{fig:prompt}.

\section{The InReAcTable System}
\label{sec:system}

\begin{figure*}[tb]
 \centering  
    \includegraphics[width=\textwidth, 
    alt={Figure 3 shows the user interface of the InReAcTable prototype system with three interconnected panels: the data subspace selection panel, the iterative exploration panel, and the insight information panel.}]{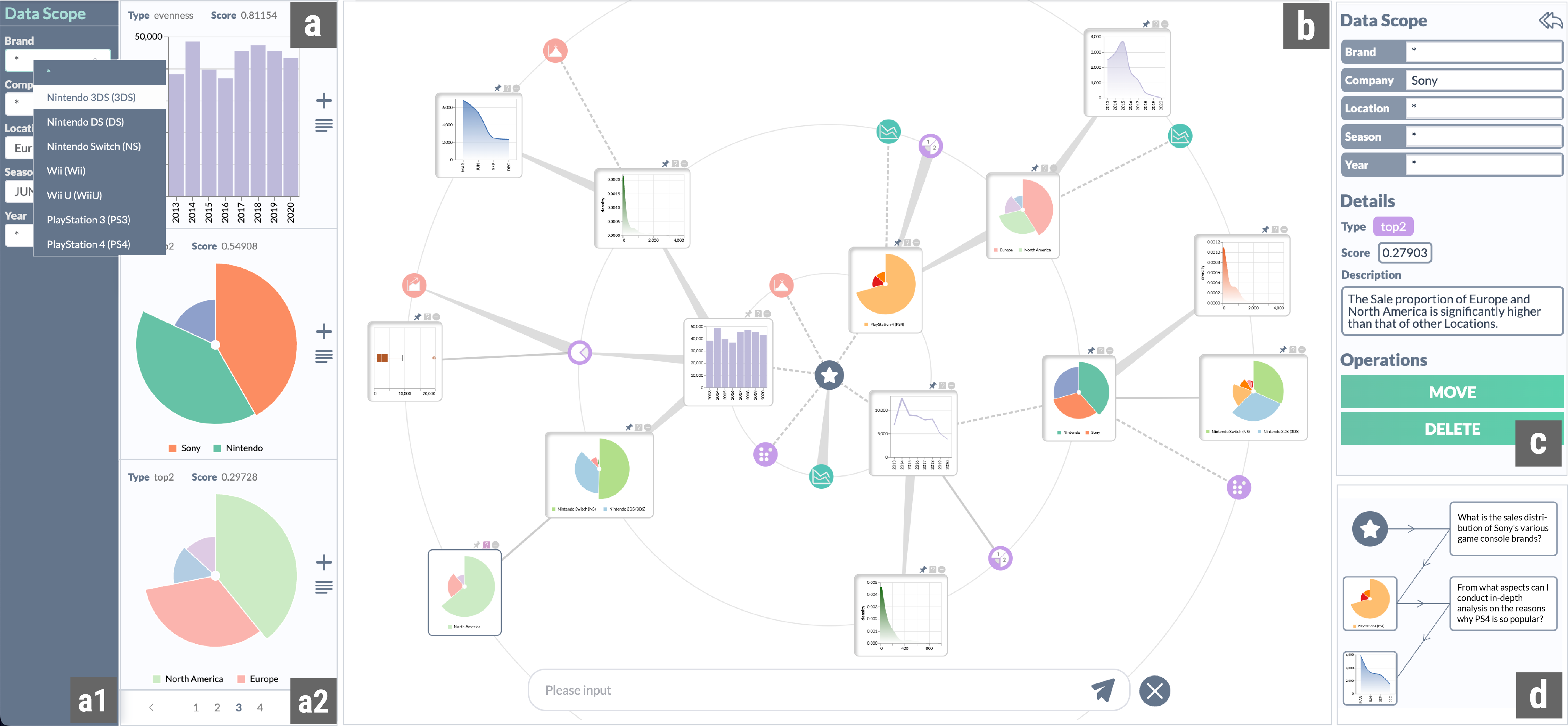}
    \caption{The user interface of the InReAcTable prototype system. In (a) data subspace selection panel, users can determine a specific data subspace by filtering attribute values in (a1), and all insights in this subspace will be displayed in the (a2) candidate insights panel. The (b) iterative exploration panel uses a circular force-directed layout diagram to map nodes and edges in the exploration path tree. Users interactively explore data through this interface. The insight information panel consists of the (c) insight detail view, presenting detailed information about the focused data insights and supporting users in modifying the exploration path, as well as the (d) exploration history view, enabling users to retrace their steps.  }
 \label{fig:system}
\end{figure*}

We develop a prototype system to assist users in visual data story construction for tabular data based on the InReAcTable framework. This section introduces the user interface and user interaction of the system.

\subsection{User Interface}
\label{sec:user-interface}

The user interface of InReAcTable system consists of three interconnected panels: the data subspace selection panel, the iterative exploration panel, and the insight information panel, as shown in Fig.~\ref{fig:system}.

\subsubsection{Data Subspace Selection Panel}
The data subspace selection panel is designed to accommodate scenarios where users have specific analytical goals in data exploration. 
In these cases, users may already have a particular focus or hypothesis, making it essential to quickly narrow down the dataset to relevant subspaces. 

As shown in Fig.~\ref{fig:system}(a1), the filter box on the left side of the panel allows users to filter data based on categorical attributes (as detailed in Sect.~\ref{sec:Abstract-Model}), 
while the insights list on the right (see Fig.~\ref{fig:system}(a2)) dynamically updates to show detailed information of all data insights extracted from the selected subspace, including visualizations, categories, types, scores, and textual descriptions. 
By separating insights according to different subspaces, this panel ensures that users can focus on the context of a particular insight, leading to more targeted exploration.

\subsubsection{Iterative Exploration Panel}

The iterative exploration panel is the core component of the InReAcTable system, 
designed to facilitate the iterative construction of visual data stories. 
To effectively visualize visual data stories, we adopt a \textbf{radial insight tree} that encodes insights as nodes and their relationships as edges, as shown in Fig.~\ref{fig:system}(b).
Although most existing systems visualize data stories as linear sequences~\cite{shi_2021_calliope, 9903555, DBLP:journals/cga/LeeRIC15} to emphasize a fixed narrative order, this sequential structure often limits flexibility in reflecting the dynamic and branching features of exploratory data analysis.

In contrast, our framework supports iterative and user-driven exploration, allowing users to flexibly shift analytical focus and redefine targets as their understanding evolves. 
To capture this non-linear progression, we represent the visual data story as a tree structure, where branches reflect alternative paths and evolving insights throughout the analysis.
We employ the radial layout for its superior spatial efficiency: it reduces visual clutter and maintains clarity even with a large number of insights. 
The tree structure not only enables users to trace their analytical path, but also supports the construction of coherent and multi-perspective data stories.

\textbf{Layout Design.}
The InReAcTable system employs a custom force-directed algorithm that is implemented using force simulation in D3.js~\cite{6064996} to dynamically arrange the nodes and edges of the graph, thus achieving a balanced layout. 
To improve clarity and usability, the nodes are constrained to concentric circles, with the radius increasing with the node's depth, visually representing the chronological stages of exploration. 
This design emphasizes the step-by-step characteristic of the exploration process and allows flexible expansion in any direction, which can optimize spatial utilization. 
A query box at the bottom of the panel enables users to input questions, with the system recommending relevant data insights for the next step.

\begin{figure}[tb]
 \centering  
    \includegraphics[width=\columnwidth,
    alt={Figure 4 shows the visual mappings of the radial insight graph in the InReAcTable system, where node colors encode categories, icons indicate insight types, and edge patterns and thicknesses represent different relation types.}]{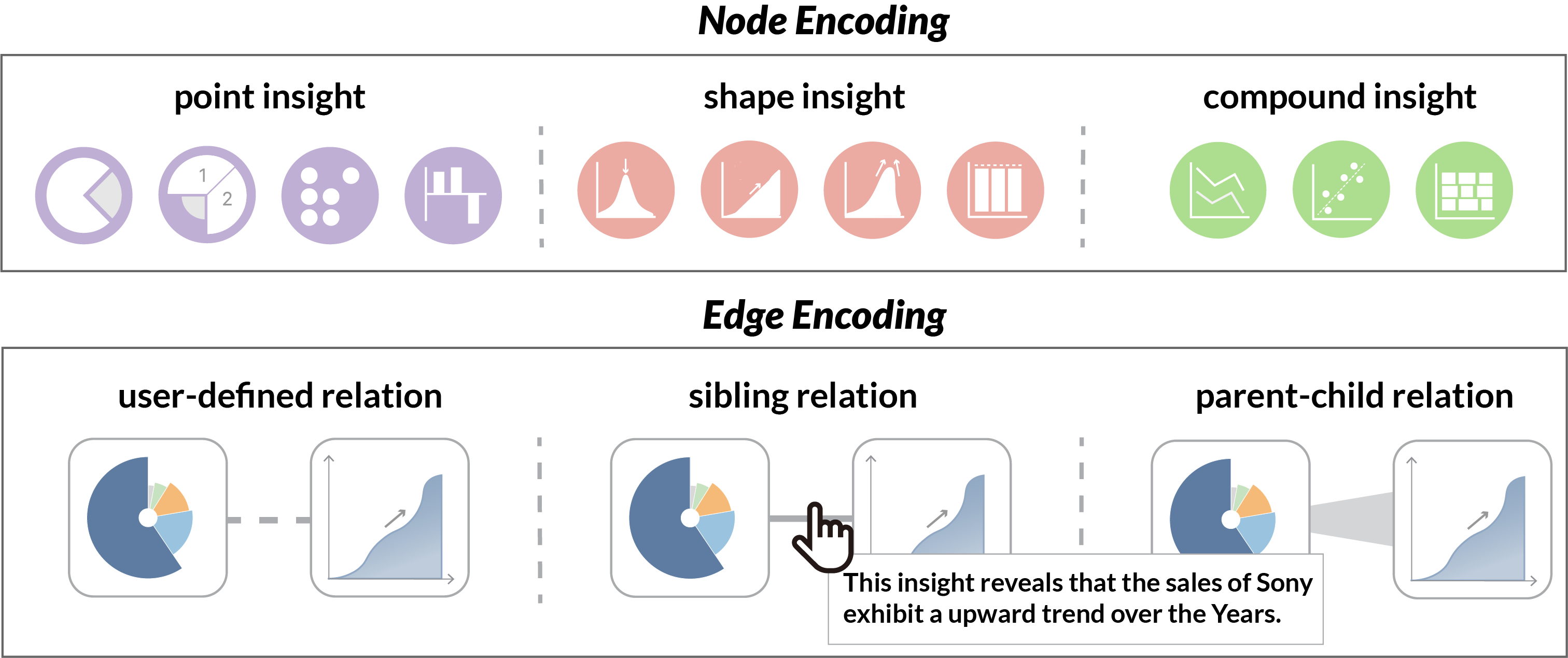}
    \caption{        
        Visual mappings of the radial insight graph in the InReAcTable system. 
        For node visual mapping, the color encodes the category, and the icons within the nodes indicate the specific types of insights. 
        For edge visual mapping, we use different patterns and thicknesses to map the three types of relations.
        Semantic connections are displayed by hovering over the edge.
    }
 \label{fig:insight_relations}
\end{figure}

\textbf{Visual Mapping.}
To reduce cognitive load and improve clarity in visualizing these insights and their interconnections, the InReAcTable system employs intuitive visual mappings for nodes and edges based on the insight categories and types defined in Sect.~\ref{sec:interaction} 
and structural relationship types defined in Sect.~\ref{sec:acting}, as shown in Fig.~\ref{fig:insight_relations}.
\begin{itemize}[leftmargin=*]

\item \textbf{Nodes.}
The color of the nodes encodes the category of the corresponding data insight, and the glyph within the node indicates its specific types.
The attribute information is displayed by hovering over the data in the charts.
During exploration, InReAcTable automatically tracks user interactions, highlighting the most recently clicked insight chart with a thicker border to denote it as the focused state. 
The system dynamically updates the target node for the query box and the insight information panel, ensuring that the interface reflects the user's current focus.

\item \textbf{Edges.}
The structural relations between insights are encoded into different styles of the edges, using solid and dashed lines, as well as varying thicknesses at different ends to represent different relations. 
Each edge's thickness gradient indicates the parent-child relationship: the thick end connects to the data insight from a larger subspace (i.e., parent), while the thin end connects to insight from a smaller subspace (i.e., child).
Uniformly thick edges link the sibling nodes, and dashed lines connect user-added nodes.
Additionally, if two nodes are connected based on LLMs' inference, the edge between them also corresponds to a semantic relation text segment.
To maintain simplicity and clarity, the system does not display the text explicitly, but allows interactive viewing of the semantic relation by hovering over the edge.
\end{itemize}

\subsubsection{Insight Information Panel} 
The insight information panel complements the iterative exploration panel by providing in-depth details about insights. 
It consists of two dynamically updated views: the insight detail view and the exploration history view.

The insight detail view (see Fig.~\ref{fig:system}(c)) offers comprehensive information about the selected node, including options to move or delete it, 
allowing users to check details and modify the exploration path flexibly. 
To facilitate context exploration, users can click the inspect button in the upper right corner of this view to apply the current node's filtering criteria directly to the data subspace filter panel, enabling quick access to all data insights within the same subspace.

The exploration history view (see Fig.~\ref{fig:system}(d)) visualizes the exploration path of the selected node, allowing users to trace their steps.
The left side displays the insight visualizations, while the right side shows the user query at each step for a quick review. 
The corresponding node in the iterative exploration panel is highlighted, enabling users to quickly pinpoint its position within the global insight tree and better comprehend the overall narrative.

\subsection{User Interaction}
\label{sec:user-interaction}

After uploading a tabular dataset, InReAcTable initially partitions the table into subspaces and performs insight extraction.
In the data subspace selection panel, users can determine a specific data subspace by filtering attribute values and extract the underlying insights.

The system displays highly significant insights in the first layer of the radial insight tree in the iterative exploration panel to inspire user exploration.
The system provides collapsed and expanded states for the insights in the radial insight tree. 
All newly added nodes are in the collapsed state, and the corresponding data insight types, such as trend and outlier, are mapped with icons to realize efficient use of screen space.
When the user clicks on a collapsed node, the node will expand into a rectangular card, showing a detailed visualization of the corresponding insight.

In the expanded state, the InReAcTable system provides three operations for insights: pin, query, and collapse.
The pin operation detaches the node from the force-directed graph layout, allowing its position to be changed only by user drag-and-drop interaction, with a shadow effect on the border indicating this fixed state, as shown in Fig.~\ref{fig:system}(b).
The query operation sets the target as the current focused node and enables users to input questions in textual format.
The collapse operation changes the expanded insight chart to its collapsed state.
Due to space constraints, users can interactively view underlying data by hovering over the insight chart or switching to the focused state to view detailed information in the insight information panel.

For insights of interest, users can view detailed information and trace their historical analysis paths in the insight information panel, particularly when users need to revisit or compare previously explored insights.
In addition, users are allowed to interactively modify the data story by rearranging branches in the radial insight tree, thereby integrating multiple insights into a coherent analytical narrative.
During this process, users can pose natural language questions in the query box and get recommended insights in the next layer of the radial insight graph under the guidance of the Action and Reasoning module, thereby iteratively constructing visual data stories.

\section{Evaluation}
\label{sec:evaluation}
To demonstrate the effectiveness of the InReAcTable framework and the usability of the InReAcTable system, we presented a real-world use case with a data analyst and conducted a user experiment compared to existing data analysis systems.

\subsection{Use Case}
\label{sec:case-study}

\begin{figure*}[tb]
 \centering  
    \includegraphics[width=\textwidth,
    alt={Figure 5 illustrates a use case of the InReAcTable system, demonstrating iterative exploration and reasoning of temporal outliers in Microsoft sales data, with a tree-structured data story capturing insights and their relationships from different perspectives.}]{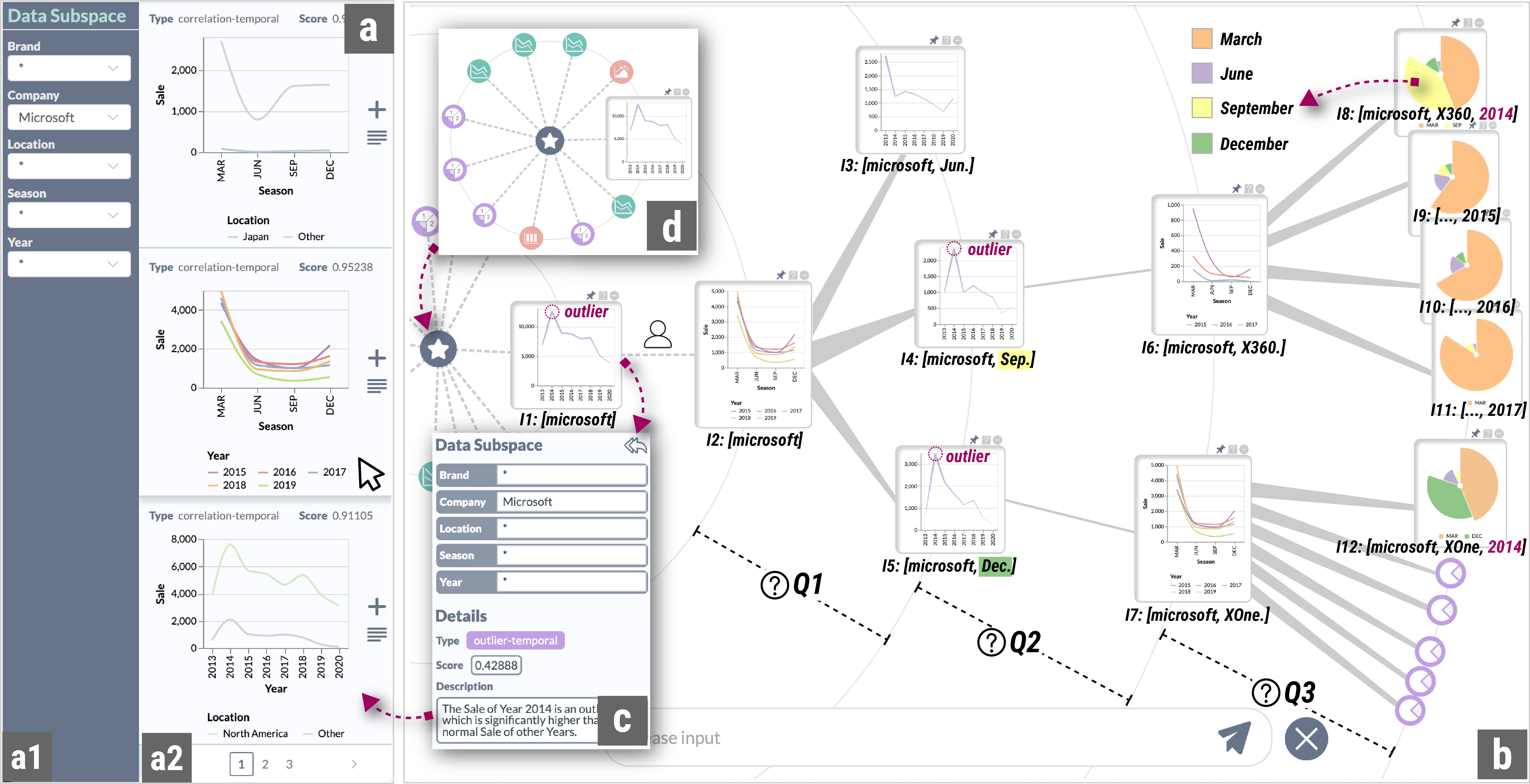}
    \caption{
        A use case demonstrating the data story construction process using the InReAcTable system, showcasing the iterative exploration and reasoning of the temporal outlier in Microsoft's sales data (I1) from different perspectives.
        The generated data story is presented in (b) as a tree-structured data narrative that captures the insights and their relationships.
    }
 \label{fig:usecase}
\end{figure*}   

We demonstrate the practical application of InReAcTable through a use case involving a data analyst with more than five years of experience in market data analysis. 
The analyst works with a real-world dataset containing sales for game consoles from various hardware companies, covering different locations and quarters from 2013 to 2020, totaling 1280 sales values. 
The dataset consists of six columns: \textit{Company}, \textit{Brand}, \textit{Location}, \textit{Season}, \textit{Year}, and \textit{Sales}, where the first five columns are categorical attributes, and \textit{Sales} is a numerical attribute. 
The Reasoning module uses OpenAI’s gpt-4-turbo model for implementation.

After uploading the sales dataset, InReAcTable first partitions the tabular data into subspaces and extracts data insights from each. 
Since the analyst does not have a specific goal, InReAcTable recommends several data insights to prompt the initial exploration, as shown in Fig.~\ref{fig:usecase}(d). 
These insights come from two sources. First, to provide an overview of the entire dataset, InReAcTable regards the whole table as a single subspace and recommends typical insights within this global space. 
Second, to highlight the most representative data insights, the system presents a set of insights with high scores from all subspaces on the first layer of the iterative exploration panel. 

After interactively checking the recommended data insights, the analyst noticed a time-series outlier insight, as shown in Fig.~\ref{fig:usecase}(I1). 
This insight, derived from the subspace \textit{S}: \textit{Company} = \textit{Microsoft}, revealed a significant sales outlier for Microsoft in 2014, followed by a continuous decline.
Intrigued by this outlier, the analyst decided to investigate its underlying causes. To do so, the analyst used the inspect button on the insight detail view (Fig.~\ref{fig:usecase}(c)) to apply this data scope to the filter panel, allowing a review of other insights within the same subspace, as shown in Fig.~\ref{fig:usecase}(a).
The analyst observed that Microsoft sales were unevenly distributed seasonally, as shown in the second insight of Fig.~\ref{fig:usecase}(a2). 
The temporal correlation data insight indicated that Microsoft generally exhibited similar seasonal trends throughout most years. 
However, the critical year of 2014, marked by abnormal sales figures, did not follow this pattern.
Considering that 2014 was an outlier in both insights, these patterns suggested that temporal factors might be a critical aspect. 
To further investigate how these factors contributed to the temporal outlier, the analyst added this insight to the next layer of the exploration path, as illustrated in Fig.~\ref{fig:usecase}(I2).

To explore the reason for the temporal outlier, the analyst posed the first question (see Q1). 
InReAcTable returned three insights detailing Microsoft's sales patterns across different months, as illustrated in Fig.~\ref{fig:usecase}b(I3, I4, I5), aligning well with the analyst's exploration objectives.
By sequentially checking the data insights, the analyst find that temporal outliers in sales exist in September (Fig.~\ref{fig:usecase}b(I4)) and December (Fig.~\ref{fig:usecase}b(I5)), which is consistent with the overall trend of Microsoft. 
The analyst speculated that the sales surge in these two months was the primary contributor to the temporal outlier.

\graybox{
Q1: I noticed a significant temporal outlier in 2014 from the sales of Microsoft and the temporal pattern in 2014 of Microsoft is quite different from other years. Are these insights correlated?
}

To explore the reasons behind the sales outliers in September and December of 2014, the analyst posed the second question (see Q2). 
InReAcTable returned two temporal correlation insights, which are the compound insights of Microsoft’s two console brands, X360 and XOne, as shown in Fig.~\ref{fig:usecase}b(I6) and Fig.~\ref{fig:usecase}b(I7).
These two insights revealed that X360 and XOne exhibited similar seasonal trends in most years except 2014. 
This suggested that the two products showed different sales patterns in 2014, implying that product-specific factors could explain the sales outlier.

\graybox{
Q2: What might be the underlying reason for Microsoft's sales surge in September and December of 2014?
}

To gain further clarification, the analyst asked the InReAcTable system to focus on different years for these two products. 
Specifically, the analyst posed the third question (see Q3) for each insight shown in Fig.~\ref{fig:usecase}b(I6, I7).
InReAcTable recommended multiple point insights focusing on different years.
From the recommended visualizations, the analyst identified that while X360's sales were generally dominated by March (Fig~\ref{fig:usecase}b(I9-I11)), there were distinct patterns in 2014.
More specifically, X360's sales in September were nearly equal to those in March, as shown in Fig.~\ref{fig:usecase}b(I8).
Similarly, for XOne in 2014, sales in December were also close to March, as shown in Fig.~\ref{fig:usecase}b(I12). 
These insights led the analyst to conclude that the sales of X360 in September and XOne in December of 2014 were the main factors that resulted in the overall sales outlier for Microsoft in that year.

\graybox{
Q3: What are the differences in sales of product X360 and XOne in 2014 among different years?
}

Through this analysis, the analyst constructed a visual data story, beginning from the temporal outlier insight corresponding to Fig.~\ref{fig:usecase}b(I1).
Initially, the analyst interactively detected a temporal outlier in Microsoft over the years from the preliminary visualization results of the dataset. 
Leveraging the \textit{\textbf{Acting}} and \textit{\textbf{Reasoning}} modules iteratively, the analyst explored various potential causes of this trend from multiple perspectives. 
This iterative exploration results in a tree-structured data narrative, capturing these insights and their relationships, as shown in Fig.~\ref{fig:usecase}(b).

\subsection{User Experiment}
We conducted a comparative user experiment to evaluate the effectiveness of the InReAcTable system.

\add{Since our system is built on the integrated InReAcTable framework, removing any module from the framework would apparently degrade the effectiveness and user experience. 
Specifically, removal of the Acting module will introduce substantial noise into the LLMs input, significantly degrading the quality of recommendation results, as demonstrated by existing works~\cite{DBLP:conf/nips/ZhouTZLW024, DBLP:conf/icml/ShiCMSDCSZ23}. 
Additionally, omitting the Reasoning module forces users to manually select data insights without guidance by browsing lots of candidates, which is a tedious and knowledge-intensive task that diminishes usability.
Given the above considerations, instead of module ablations, we chose to compare the InReAcTable system against two widely used existing systems to provide stronger external validity.
}

Due to the public availability of some previous systems, we selected QuickSight\footnote{https://aws.amazon.com/cn/quicksight} as a representative example of automated generation tools and vizGPT\footnote{https://vizgpt.ai/} as an interactive tool driven by LLMs to ensure a comparison between two distinct categories of systems. 
This comparison allowed us to benchmark our system against both a widely adopted automated solution and a mature LLM-based method. 

\subsubsection{Experimental Setup}

\textbf{Participants. }
This experiment recruited 18 participants (5 females and 13 males, aged 19-27) to evaluate the practicality and effectiveness of the InReAcTable system in data story construction. 
The participants included undergraduate and graduate students from diverse academic backgrounds, including computer science, economics, and mathematics.
All participants had a basic understanding of data analysis and had experience using basic tools such as Excel and Google Sheets for statistical analysis.
Some participants (n=7) were familiar with professional data analysis tools (e.g., PowerBI~\cite{ding_2019_quickinsights} and Tableau~\cite{tableau2024}) and had accumulated experience in complex data analysis.
Furthermore, all participants had experience applying generative AI tools (e.g., ChatGPT~\cite{openai2023chatgpt}) to assist in their analysis within real-world projects.

\textbf{Dataset and Apparatus. }
In this study, all participants used the same dataset to conduct the experiments, thereby avoiding unnecessary confounding variables due to varying richness in the value of different datasets. 
The dataset used in the experiment was the sales dataset of game consoles, as mentioned in Sect.~\ref{sec:case-study}.
The experiment was conducted in a quiet and isolated laboratory to minimize external disturbances that could affect the results. 
Before the experiment began, all participants signed informed consent forms and each received compensation equivalent to \$20 after the completion of the experiment.

\subsubsection{Experimental Procedure}

\deleted{This study employed a within-subject design. }
Considering that participants might develop prior knowledge of the dataset after using one tool, potentially influencing their performance with subsequent tools, we adopted a Latin square design to counterbalance the order of system usage at the group level. 
Specifically, the three systems were arranged into six sequences using the Latin square design. 
The 18 participants were randomly assigned to six groups of three. 
Each of the six experimental sessions was conducted independently and lasted approximately 1.5 hours.
\add{The potential impact of ordering effects at the individual level is assessed, and the results are reported in Sect.~\ref{sec:Results_analysis}.}

\textbf{Training (20 minutes).}
At the start of the experiment, we outlined the experimental procedure and provided relevant background knowledge, including the dataset source and key concepts. 
The tutorials for the three systems demonstrated the functions of each module, followed by hands-on practice to address any difficulties. 
We introduced the specific task requirements once all participants were familiar with the systems. 
The training stage was controlled to be within 30 minutes.

\textbf{Task (70 minutes).}
After the training stage, participants were asked to independently complete two experimental tasks sequentially.
The inclusion of these two tasks in the study was informed by Sect.~\ref{sec:preliminary_study}, which revealed that user exploration typically falls into two distinct modes: 
\begin{itemize}[leftmargin=*]
\item \textbf{Goal-seeking exploration mode}, where users have an unclear query and rely on the system to provide high-scoring and diverse insights to inspire exploration; 
\item \textbf{Targeted exploration mode}, where users possess a clear query and expect the system to recommend insights that align closely with their intent. 
\end{itemize}
These two tasks were designed to evaluate the system’s ability to support both modes, focusing on the balance between fostering creativity (divergent thinking) and delivering precision (convergent thinking).

\begin{itemize}[leftmargin=*]
    \item \textbf{Task 1: Freeform table exploration.}
    \add{This task adopted a within-subjects design. }
    Participants were instructed to use three different systems to explore the dataset freely, dedicating 20 minutes to each system. 
    The primary goal of this task was to identify and organize valuable insights into data stories.
    \deleted{During the process, all activities were recorded using screen capture technology. Participants were required to take screenshots of the discovered data stories and append necessary textual descriptions, documenting the relations between connected insights and their analytical ideas.}
    Participants were explicitly informed in advance that both the quantity and quality (logicality, coherence, diversity, and relevance, see Sect.~\ref{sec:Results_analysis}) of the data stories were important.
    \deleted{After completing Task 1, all participants gained a thorough understanding of all three systems. This ensured that the subsequent task was fair and not influenced by the order in which the systems were used in Task 1.}
    
    \item \textbf{Task 2: Target-driven table exploration.} 
    \add{This task followed a between-subjects design.}
    In Task 2, 
    participants were randomly assigned to one of the three systems, with equal distribution. 
    All participants were given the same initial insight and a specific question. 
    They had 10 minutes to explore related insights and create a data story to answer the question. 
    \deleted{The recording requirements were identical to Task 1.} 
    The task design ensured that participants' performance in Task 2 was based on their understanding and capability with the assigned system rather than any potential bias from the sequence of system usage in Task 1.
\end{itemize}

\add{During both tasks, all participant activities were recorded using screen capture applications. 
Participants were required to take screenshots of the discovered data stories and append necessary textual descriptions, documenting the relations between connected insights and their analytical ideas.
Results from the two tasks were collected and analyzed separately.}

\textbf{Interview (10 minutes).}
During the experiment, participants were encouraged to engage actively and ask questions, with all feedback meticulously recorded. 
We used a five-point Likert scale to collect participants' evaluations of each system and their analysis experience.
After the experiment, we conducted a group discussion (5 minutes) and individual interviews (5 minutes) to discuss their feedback and subjective experiences during the data story construction. 
\add{The group discussion was conducted in a co-located workshop setting, allowing participants to exchange thoughts and reflect on their overall experience.}

\subsubsection{Results and Analysis}
\label{sec:Results_analysis}

\add{We first assessed the potential impact of ordering effects on the individual level. 
We analyze the outputs of the participants and find that they did not construct identical stories across systems, and there were few overlapping insights. 
This can be attributed to the substantial volume of available insights in the dataset (over 3,000) and the distinct modes of exploration across systems, which guided participants along different analytical paths. 
Therefore, individual-level ordering effects appear to be minimal.}

Subsequently, \add{the experiment results of Task 1 and Task 2 are analyzed separately from three perspectives,} quantitative evaluation, qualitative evaluation, and user feedback, to comprehensively assess the effectiveness and usability of the system.

\begin{figure}[tb]
 \centering  
    \includegraphics[width=\columnwidth,
    alt={Figure 6 shows bar charts comparing the number of data stories, the number of insights per story, and the total number of insights generated by participants using InReAcTable with baseline systems, with InReAcTable outperforming the baseline systems in all three metrics.}]{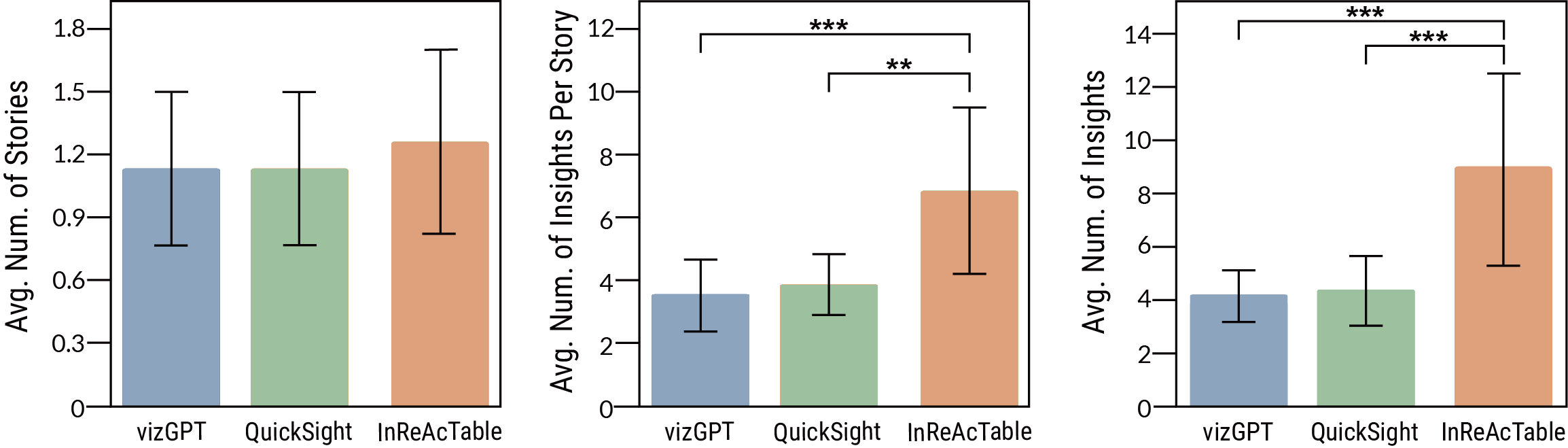}
    \caption{Quantitative evaluation results of user experiments. 
    Comparison of the number of data stories (left), the number of insights per story (middle), and the total number of insights (right) generated by participants using InReAcTable and baseline systems in Task 1. 
    (** \textit{p} < 0.01, *** \textit{p} < 0.001)}
 \label{fig:quantitative-evaluation-results}
\end{figure}  

\textbf{Quantitative evaluation.}
\add{We first performed the Shapiro-Wilk Test to verify the normality of the data distributions and then conducted one-way ANOVA to examine differences across three systems.
The ANOVA revealed significant differences ($p < 0.05$) for two quantitative metrics: the number of insights per story and the total number of insights.}
Post-hoc paired t-tests with Bonferroni correction ($\alpha = 0.025$) were then applied to compare InReAcTable with each baseline individually, 
which showed that participants using InReAcTable produced significantly more insights per story ($p < 0.01$) and a higher total number of insights ($p < 0.01$), suggesting that InReAcTable is more effective in supporting users to uncover and synthesize data insights for data story construction. 
\add{For the metric number of data stories, the ANOVA results indicate no statistically significant difference across systems, thus no post-hoc tests were conducted for this metric.}
\deleted{For quantitative evaluation, paired t-tests were performed on the number of data stories, the number of insights per story, and the total number of insights, comparing InReAcTable with baseline systems to determine statistically significant differences in participant performance.}
The results of the quantitative evaluation are shown in Fig.~\ref{fig:quantitative-evaluation-results}.

In addition, we observed that InReAcTable users tended to construct profound data stories with multiple insights. 
The hierarchical structure of the story tree effectively represents the analysis structure, with each subtree representing a specific direction and multiple subtrees combined to support an analytical topic from various perspectives. 
In contrast, participants using vizGPT and QuickSight were inclined to create shorter stories with fewer data insights. 
This might be due to the limited understanding of tabular data and the linear narrative style, which resulted in simpler and less extensive stories.

\begin{figure}[tb]
 \centering  
    \includegraphics[width=\columnwidth,
    alt={Figure 7 presents bar charts comparing expert ratings on logicality, coherence, diversity, and relevance of data stories constructed using InReAcTable and baseline systems in Task 1 and Task 2, demonstrating that InReAcTable achieves higher scores on all metrics.}]{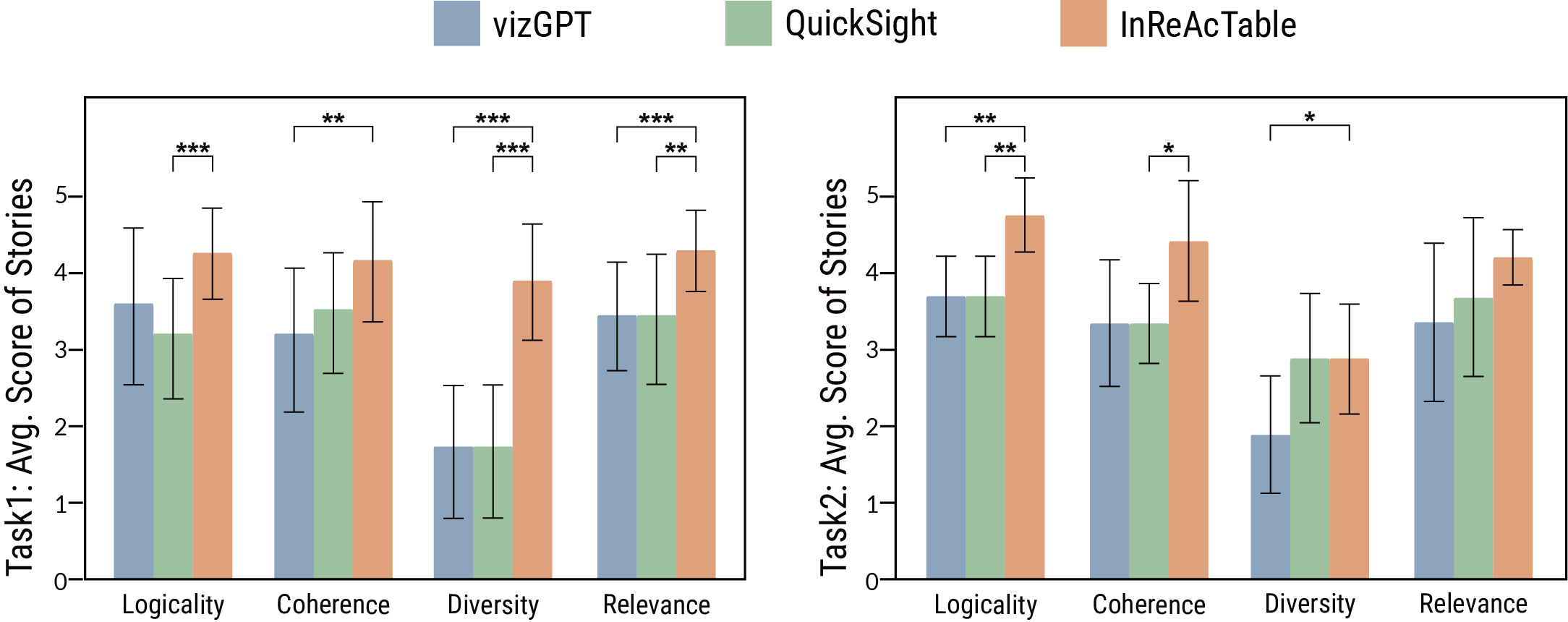}
    \caption{Qualitative evaluation results of data stories constructed using InReAcTable and other baseline systems.
    Comparison of the expert ratings on logicality, coherence, diversity, and relevance in Task 1 (left) and Task 2 (right).\\ 
    (* \textit{p} < 0.05, ** \textit{p} < 0.01, *** \textit{p} < 0.001)}
    
 \label{fig:qualitative-evaluation-results}
\end{figure} 

\textbf{Qualitative evaluation.}
For the qualitative evaluation, we invited five experts in visual data storytelling to assess the quality of the data stories constructed by the participants. 
Each expert has more than six years of field research experience and extensive practical experience in creating and evaluating high-quality data stories.
It is important to note that the experts were unaware which stories were generated using our system during the evaluation.
The experts rated the data stories from Task 1 and Task 2 using a five-point Likert scale across four dimensions: logicality, coherence, diversity, and relevance. 
The results are presented in Fig.~\ref{fig:qualitative-evaluation-results}.

\begin{itemize}[leftmargin=*]
    \item \textbf{Logicality.} 
    According to the expert evaluations, 
    nearly all participants (n=16) using InReAcTable produced logically consistent stories with strong connections between insights. 
    Similarly, more than half of the stories generated using vizGPT (n=12), the LLM-driven insight recommendation system, were also deemed logical.  
    This indicates how the enhanced reasoning capabilities of the LLMs supported users in maintaining a rigorous thought process with minimal logical flaws. 
    In contrast, most QuickSight users (n=13) received scores of 3 or less, reflecting the difficulty of relying solely on human interpretation of the charts to determine the next steps in construction.
    
    \item \textbf{Coherence.}  
    In Task 1, there were no significant differences in coherence across the stories generated using vizGPT (n=7), QuickSight (n=11), and InReAcTable (n=13), as participants followed relatively clear exploration paths under freeform conditions. 
    However, in Task 2, 
    more than half of the stories created using vizGPT and QuickSight were rated as neutral or lacking coherence, while only one data story from a InReAcTable user was rated as neutral.
    This indicates that once an analysis starting point is determined, InReAcTable can more effectively build structural and semantic relations between insights.
    Consequently, it provides users with a clear direction, enhancing the coherence and effectiveness of the construction process.
    
    \item \textbf{Diversity.} 
    In Task 1, experts pointed out that 
    users of vizGPT and QuickSight performed poorly in diversity, with only one user from each system achieving a score indicating diversity.
    These results suggest that during the freeform exploration, the single-turn dialogue model vizGPT tends to recommend similar insights, leading to repetitive analytical paths and homogeneous data stories. 
    QuickSight also exhibited low diversity, possibly due to users' tendency to rely on accustomed types of visualizations. 
    In contrast, data stories generated with InReAcTable demonstrated a greater variety of perspectives and novel discoveries. 
    Different users explored the data from various aspects, covering various data types and visualization forms, leading to rich and diverse expressions. 

    \item \textbf{Relevance.} 
    Nearly all data stories generated by InReAcTable users (n=17, with 6 rated as strongly relevant and 11 as relevant) were considered to be closely related to the analysis topic.
    The insights included in these stories directly supported or answered the users' key questions, avoiding any deviation from the topic or inclusion of irrelevant information.    
\end{itemize}

\begin{figure}[tb]
 \centering  
    \includegraphics[width=\columnwidth,
    alt={Figure 8 shows histograms comparing participant feedback on system design across vizGPT, QuickSight, and InReAcTable, indicating that participants are highly satisfied with the system interface of InReAcTable.}]{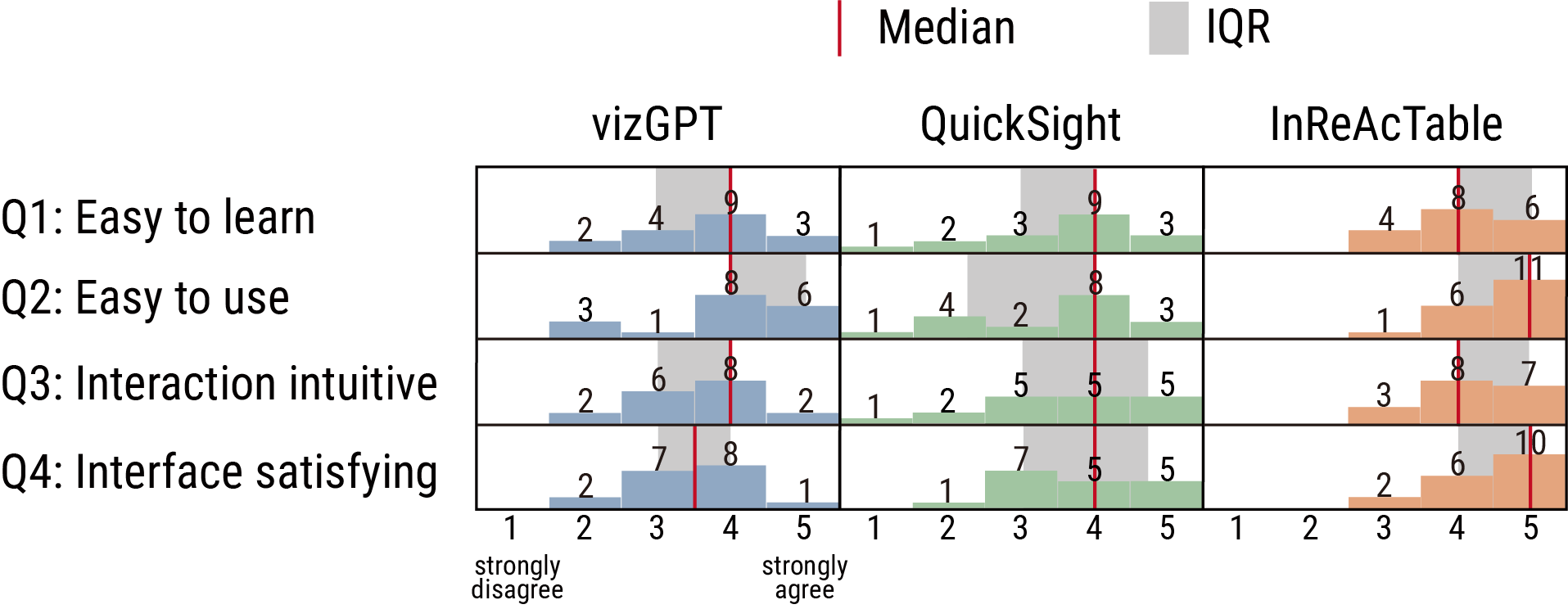}
    \caption{
    Comparison of the participant feedback on system design across vizGPT, QuickSight, and InReAcTable. 
    The result indicates that participants are highly satisfied with the system interface of InReAcTable.
    }
 \label{fig:feedback1}
\end{figure}

\begin{figure}[tb]
 \centering  
    \includegraphics[width=\columnwidth,
    alt={Figure 9 shows histograms comparing participant feedback on the data story construction experience across vizGPT, QuickSight, and InReAcTable. The high median scores and narrow interquartile range indicate uniform and positive feedback for InReAcTable.}]{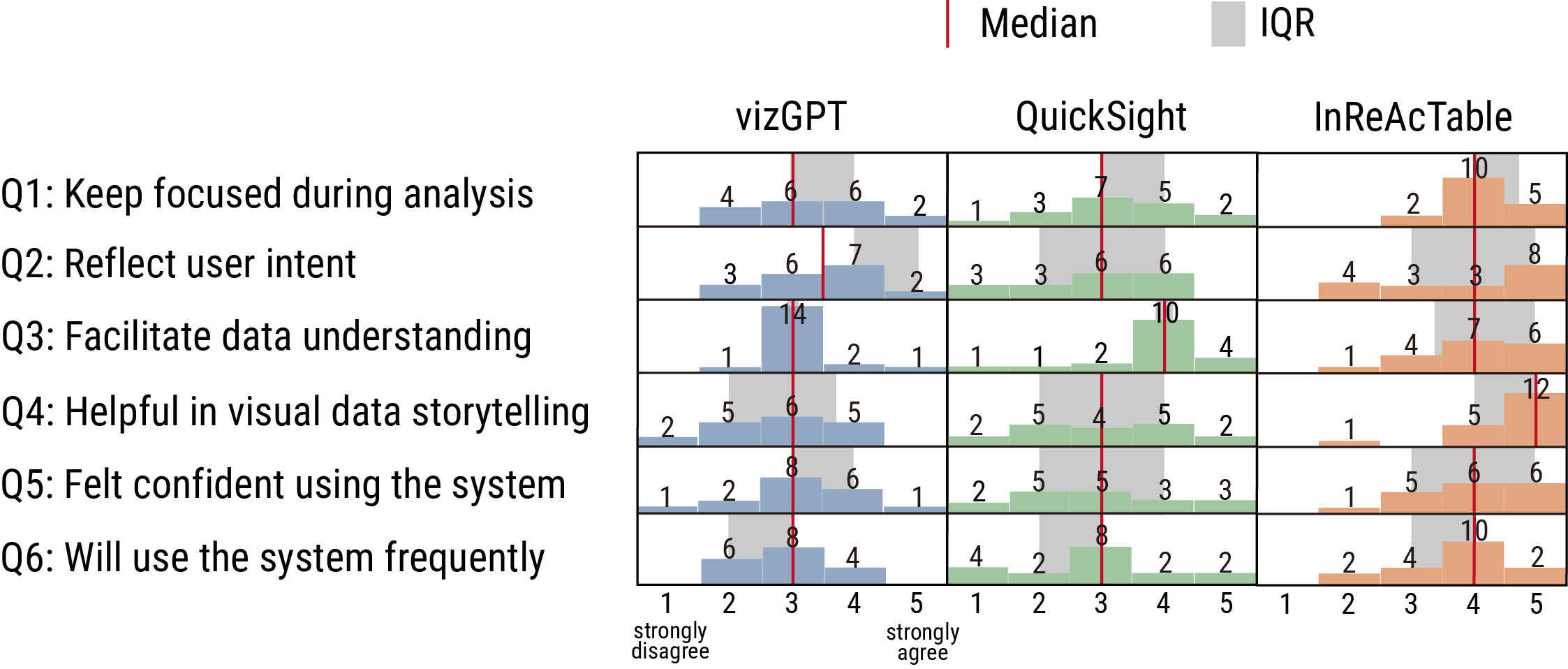}
    \caption{
    The result of the participant feedback on the data story construction experience. 
    The high median scores and narrow interquartile range indicate uniform and positive feedback for InReAcTable.
    }
 \label{fig:feedback2}
\end{figure}

\textbf{Participant feedback.}
The user feedback on both the system design and the data story construction experience was overwhelmingly positive, as illustrated in Fig.~\ref{fig:feedback1} and Fig.~\ref{fig:feedback2}.
The results are presented in histograms according to the principles outlined by Dragicevic et al.~\cite{dragicevic2016fair}.

\textbf{Feedback on system design.}
Participants rated the InReAcTable system highly on ease of learning (Q1), ease of use (Q2), intuitive interaction (Q3), and interface satisfaction (Q4). 
The median scores for InReAcTable were consistently at the higher end of the scale, with most participants agreeing with the positive statements. 

The results show that the dialogue-based interface of vizGPT facilitates usage (Q1 and Q2) but has limited functionality and a simple interaction mode (Q3 and Q4).
In contrast, QuickSight, a feature-rich commercial tool, scored higher on functionality-related aspects (Q3 and Q4) but faced challenges with complexity and a steep learning curve (Q1 and Q2).
The InReAcTable system integrates user-friendly design and essential analytical capabilities, balancing simplicity and functionality.
During interviews, many participants acknowledged the system's effective and flexible interaction in constructing data stories. 
One participant noted, ``\textit{I could obtain the insights I was interested in with just a few clicks on the nodes and by posing a simple question. The system's intuitive and visually appealing mapping meant I hardly needed to do any extra thinking.}''

\textbf{Feedback on data story construction experience.}
In terms of data story construction experience, InReAcTable also received positive feedback.
Most participants indicated that it helped them stay focused (Q1), accurately reflected their intent (Q2), facilitated data understanding (Q3), and supported visual storytelling (Q4). 
Additionally, participants expressed high confidence in using the InReAcTable system (Q5) and a strong willingness to use it frequently (Q6). 

The feedback of vizGPT suggests that although it supports users in expressing intent (Q2),
the homogeneity of the LLM-driven recommendations often restricts users to a limited set of questions. 
This reduces the ability to uncover diverse insights (Q4) and leads to a narrow understanding of the data (Q3). 
On the other hand, QuickSight offers advanced operations that enhance data understanding (Q3) and visual storytelling (Q4). 
However, it lacks guidance for users' analytical directions (Q2), often leading to confusion (Q1) and reduced user confidence (Q5).
InReAcTable balances insight recommendations with prompt responsiveness to user intent.
Participants reported that it often broadened their analytical perspective, allowing them to step out of their current focus on a specific chart to explore other related insights. 
One participant remarked, ``\textit{This guidance was particularly valuable when my analysis goals were unclear}.''
Another participant appreciated the LLMs’ recommendations and was surprised that the reasons were sometimes directly pointed to a particular value of the insight.
He mentioned, ``\textit{I can express my intent to InReAcTable and receive hints, which helps me decide whether to continue with a particular line of inquiry.}''

\add{
During the group discussion, participants expressed satisfaction with the quality of the system’s recommendations, especially given the size and complexity of the datasets used. 
One participant noted, ``\textit{Based on my previous experience with LLMs, I thought it would generate vague and repetitive answers, even hallucinations, but InReAcTable gave surprisingly concrete and varied suggestions.}''
These reflections highlight the impact of the Acting module, which structures context-aware prompts to filter noise and focus the LLM input, effectively preprocessing the data before generating recommendations.

In addition, participants also emphasized the value of the Reasoning module, describing it as offering ``just enough push'' to help them move forward without being overwhelmed by irrelevant information. 
Several participants reported experiencing ``aha moments'' when the system provided explanations that surfaced unexpected connections within the data. 
One participant remarked, ``\textit{It sometimes points out patterns that I hadn’t noticed at all---those moments really inspire me to rethink my assumptions.}''
These reflections suggest that the Reasoning module not only prevents analytical stagnation but also enhances users' trust by making the system's suggestions more transparent and cognitively meaningful.
}

\section{Discussion and Future Work}

\add{
\textbf{Generalizability of the InReAcTable framework.} 
The InReAcTable framework investigates how extracted data insights and their interrelationships can be leveraged for exploratory visual analysis through integration with LLMs. This paradigm is adaptable to diverse types of data and analytical tasks.

At the data level, although our work focuses on tabular data, the insight-centric approach, emphasizing the extraction and linking of insights, can be extended to other data types such as time series data~\cite{DBLP:journals/tvcg/ShiCCJXJGC24}, hierarchical data~\cite{treevis2011, li2023GoTreeScape, li2020gotree, li2020barcodetree}, and multimodal data~\cite{DBLP:conf/icmi/TurbevilleMSMPL24}. This adaptability indicates that insight-based frameworks can support exploratory analysis across a wide range of domains, enabling more flexible and scalable data exploration.

At the task level, data story construction can be extended to the broader field of visual analysis.
Although many LLM-powered visual analytics systems~\cite{leva2024, zhao2024lightvalightweightvisualanalytics} rely on prompt-driven interactions, our approach emphasizes the benefits of modular reasoning and acting architectures in managing the complexity of visual analysis tasks.
This highlights new opportunities for the design of future visual analytics systems, where flexibility and modularity can be used to create more adaptive and user-centric tools.}

\textbf{Scalability of the InReAcTable framework.}
Although InReAcTable currently supports various categories of insights, such as point, shape, and compound insights, its ability to handle increasingly complex datasets can be further enhanced by expanding the range of available types of insight. 

In addition, InReAcTable could benefit from extending the set of predefined logical relations between insights to further improve scalability. 
Although these relations currently help users navigate between data subspaces, the inclusion of more nuanced or customizable relations would better accommodate diverse datasets. 
Furthermore, integrating machine learning techniques to dynamically infer new logical connections based on user behavior or emerging data patterns could create a more personalized and adaptive analysis experience. 

From the user interaction perspective, the requirement for more flexible filtering interactions becomes crucial. 
The current system provides basic filtering by data subspaces, but advanced capabilities such as multilevel filtering, hierarchical filtering, and filtering by specific insight types can be implemented in the future.
These enhancements would not only improve usability by allowing users to focus on the most relevant insights, but also prevent cognitive overload, facilitating more effective data exploration.

\textbf{Accuracy and response time of user interaction. }
The performance of LLMs within InReAcTable has been promising, but raises concerns about accuracy. Although the system leverages LLMs to generate insightful recommendations, inherent limitations such as hallucinations and biases remain. Future iterations could address these issues by incorporating more robust training data, including domain-specific datasets, to improve the accuracy of the insights generated. Furthermore, employing human-in-the-loop approaches for validation could enhance the reliability of LLM outputs. 

In addition, response time is a critical factor in maintaining user engagement, particularly in large-scale data exploration. Although LLMs can deliver accurate insights, processing time becomes a bottleneck, especially with complex queries or large datasets. 
Future work could focus on reducing response time through model distillation, which can accelerate inference without sacrificing accuracy.

\section{Conclusion}

In this work, we propose InReAcTable, a framework that seamlessly integrates structure filtering and semantic reasoning to dynamically recommend relevant insights that align with the user's analytical goals.
Based on the InReAcTable framework, we develop an interactive system to facilitate the construction of visual data stories from tabular datasets. 
The case study with a domain expert validates the system’s effectiveness and usability on real-world datasets. 
Moreover, comparative evaluations with existing solutions substantiate the framework's ability to support both freeform exploratory and target-driven data story construction. 
In the future, we plan to extend the InReAcTable framework to more diverse data analysis scenarios, aiming to create a generalized LLM-powered paradigm for visual data story construction.

\begin{acks}
We thank the anonymous reviewers for their valuable comments. This work is supported by NSFC (62302038 and U2268205), Young Elite Scientists Sponsorship Program by CAST (2023QNRC001).
\end{acks}

\bibliographystyle{ACM-Reference-Format}
\bibliography{main}

\end{document}